\documentclass[twocolumn]{aastex62}

\newcommand{\todo}{\ifmmode \text{\color{purple}\Huge{\(\bullet\)}} \else {\color{purple}{\Huge$\bullet$}}\fi}
\newcommand{\finish}{\ifmmode \text{\color{blue}\Huge{\(\bullet\)}} \else {\color{blue}{\Huge$\bullet$}}\fi}

\newcommand{\mstar}{M_{\star}}
\newcommand{\msun}{M_{\odot}}
\newcommand{\mbh}{M_{\rm BH}}

\newcommand{\um}{\mathrm{\mu m}}

\newcommand{\lbol}{L_\mathrm{AGN,bol}}
\newcommand{\ledd}{L_{\rm Edd}}

\newcommand{\R}{\mathcal{R_{\rm rest}}}

\newcommand{\lambdaedd}{\lambda_\mathrm{Edd}}

\usepackage{natbib}
\usepackage{amsmath}
\usepackage{multirow} 
\usepackage{textgreek}
\usepackage{xcolor}
\usepackage{booktabs} 
\usepackage{ifthen} 
\received{\today}
\submitjournal{ApJ}

\shorttitle{Radio-loud DOGs}
\shortauthors{Fukuchi \& Ichikawa}

\DeclareUnicodeCharacter{2212}{-} 
\begin{document}

\title{Finding of a Radio-Loud Dust Obscured Galaxies J1416+0102: a super Eddington accretion system}

\title{J1406+0102: Dust Obscured Galaxy Hiding Super Eddington Accretion System with Bright Radio Emission}

\correspondingauthor{Hikaru Fukuchi, Kohei Ichikawa}
\email{hikaru.fukuchi@astr.tohoku.ac.jp, ichikawa.waseda@gmail.com}

\author[0000-0001-7557-6854]{Hikaru Fukuchi}
\affil{Astronomical Institute, Tohoku University, Aramaki, Aoba-ku, Sendai, Miyagi 980-8578, Japan}

\author[0000-0002-4377-903X]{Kohei Ichikawa}
\affiliation{Global Center for Science and Engineering, Faculty of Science and Engineering, Waseda University, 3-4-1, Okubo, Shinjuku, Tokyo 169-8555, Japan}
\affiliation{
Department of Physics, School of Advanced Science and Engineering, Faculty of Science and Engineering, Waseda University, 3-4-1,
Okubo, Shinjuku, Tokyo 169-8555, Japan}
\affil{Frontier Research Institute for Interdisciplinary Sciences, Tohoku University, Sendai 980-8578, Japan}
\email{ichikawa.waseda@gmail.com}

\author[0000-0002-2651-1701]{Masayuki Akiyama}
\affil{Astronomical Institute, Tohoku University, Aramaki, Aoba-ku, Sendai, Miyagi 980-8578, Japan}

\author[0000-0003-2579-7266]{Shigeo S. Kimura}
\affil{Astronomical Institute, Tohoku University, Aramaki, Aoba-ku, Sendai, Miyagi 980-8578, Japan}
\affil{Frontier Research Institute for Interdisciplinary Sciences, Tohoku University, Sendai 980-8578, Japan}

\author[0000-0002-3531-7863]{Yoshiki Toba}
\affil{National Astronomical Observatory of Japan, 2-21-1 Osawa, Mitaka, Tokyo 181-8588, Japan}
\affil{Academia Sinica Institute of Astronomy and Astrophysics, 11F of Astronomy-Mathematics Building, AS/NTU, No.1, Section 4, Roosevelt Road, Taipei 10617, Taiwan}
\affil{Research Center for Space and Cosmic Evolution, Ehime University, 2-5 Bunkyo-cho, Matsuyama, Ehime 790-8577, Japan}

\author[0000-0001-9840-4959]{Kohei Inayoshi}
\affil{Kavli Institute for Astronomy and Astrophysics, Peking University, Beijing 100871, People’s Republic of China}

\author[0000-0002-5197-8944]{Akatoki Noboriguchi}
\affil{School of General Education, Shinshu University, 3-1-1 Asahi, Matsumoto, Nagano 390-8621}

\author[0000-0002-3866-9645]{Toshihiro Kawaguchi}
\affil{Department of Economics, Management and Information Science,
  Onomichi City University, Hisayamada 1600-2, Onomichi, Hiroshima
722-8506, Japan}

\author[0000-0003-2682-473X]{Xiaoyang Chen}
\affil{
Department of Physics, School of Advanced Science and Engineering, Faculty of Science and Engineering, Waseda University, 3-4-1,
Okubo, Shinjuku, Tokyo 169-8555, Japan}

\author[0000-0002-5956-8018]{Itsna K. Fitriana}
\affil{National Astronomical Observatory of Japan, 2-21-1 Osawa, Mitaka, Tokyo 181-8588, Japan}
\affil{Department of Astronomy, Institut Teknologi Bandung, Jl. Ganesha 10, Bandung 40132, Indonesia}



\begin{abstract}
Recent high-$z$ quasar observations strongly indicate that super-Eddington accretion is a crucial phase to describe the presence of supermassive black holes (SMBHs) with $M_\mathrm{BH} \gtrsim 10^9 M_\odot$ at $z \gtrsim 7$. 
Motivated by theoretical predictions that the super-Eddington phase efficiently produces outflows and jets bright in radio bands, we identify a super-Eddington radio-loud dust-obscured galaxy (DOG) J1406+0102 at $z=0.2367$. This source is discovered by cross-matching the infrared-bright DOGs of \cite{nob19} with the VLA/FIRST 1.4~GHz radio survey data and the SDSS optical spectral catalog.
J1406+0102 shows broad components in the Balmer lines, and by assuming that those lines are from the broad line region, they give BH mass estimations of $\log\ (\mbh/\msun)=7.30 \pm 0.25$. Combined with an AGN luminosity of
$\log (L_\mathrm{bol,[OIII]}/\mathrm{erg}~\mathrm{s}^{-1}) 
 = 45.91\pm0.38$ estimated from the intrinsic [O~{\sc{iii}}] luminosity, this implies a super-Eddington accretion rate of $\lambdaedd \simeq 3$.
We show that 
1) J1406+0102 shows strong AGN feedback, with the [O~{\sc{iii}}] outflow velocity exceeds the escape velocity of the host galaxy halo and a kinetic efficiency of $\approx$~8\%, sufficient to quench the host galaxy;
2) its expected growth trajectory places it on an over-massive BH track; and 3) if representative of radio-loud DOGs, such sources can contribute significantly to the high-energy ($\gtrsim$~100 TeV) cosmic neutrino background.
\end{abstract}


\section{Introduction}\label{sec:intro}
Recent surveys of high-$z$ quasars have revealed that supermassive black holes (SMBHs)
with mass of $\simeq 10^{9}\ \msun$ already exist 
 at redshift $z \gtrsim 7$
\citep[e.g.,][]{wan21}.
These sources require massive BH seeds and/or super-Eddington accretion ($\lambdaedd \equiv \lbol / \ledd  \geq 1$) to SMBHs \citep[e.g.,][]{ina20}, where $\lbol$ is AGN bolometric luminosity and $\ledd$ is Eddington luminosity ($\ledd \simeq 1.26 \times 10^{38} (\mbh/\msun)$~erg~s$^{-1}$). 
Thus, super-Eddington accreting sources are important targets for excavating detailed BH growth,
leading to intensive theoretical and observational studies~\citep{ina20,gre20}.
\begin{figure*}
    \centering
    \includegraphics[width=1.0\textwidth]{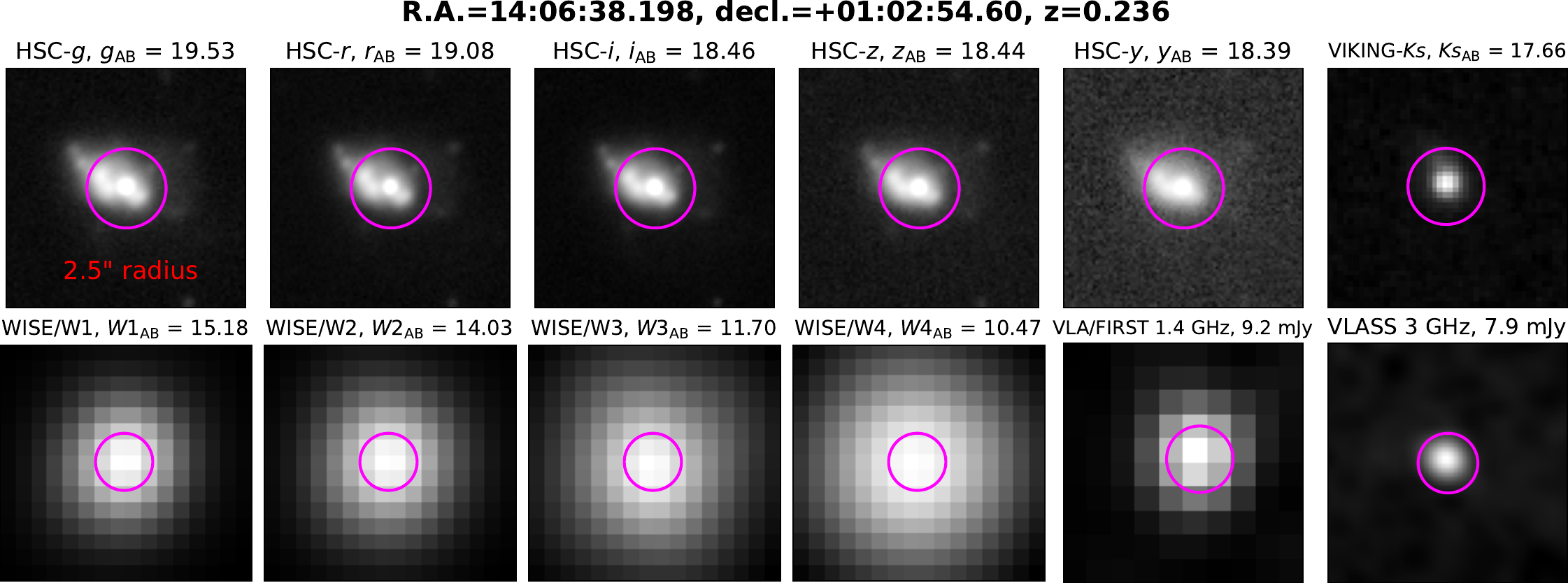}~
    \caption{Multiwavelength cutout images of J1406 15~arcsec $\times$ 15~arcsec), with the name of each photometric band and its corresponding AB magnitude(flux density for radio data) above each image. The magenta circle, with a radius of 2.5~arcsec (physical sacale of $\sim$ 4.7 kpc), represents the typical angular resolution of the VLASS survey.
  }
    \label{fig:cutout_image}
\end{figure*}
However, the currently known super-Eddington sample might be
only the tip of the iceberg since the super-Eddington phase is expected to last a short timescale of the order of 1-10~Myr based on radiative hydrodynamic simulations \citep[e.g.,][]{ina22} 
and semi-analytic model~\citep[e.g.,][]{shi20}.
In addition, most of the time they are highly obscured by the surrounding gas~\citep[e.g.,][]{hop08}.

A major merger of two gas-rich galaxies is an efficient path to trigger rapid mass accretion onto SMBHs \citep[e.g.,][]{hop06}.
They are expected to be heavily obscured by surrounding dust, which would produce extremely red color between the optical and infrared (IR) bands, and sometimes they are called dust-obscured galaxies \citep[DOGs: e.g.,][]{dey08}.
Our team recently found IR-bright DOGs
whose optical to IR color are extremely red with 
$(i - [22]) >$ 7, where $i$ and $[22]$ are HSC $i$-band and {\it WISE}~22~$\mu \mathrm{m}$ AB magnitudes \citep{tob15,tob17_clustering,nob19,nob25,yos25} by using Subaru/Hyper Suprime-Cam (HSC:~\citealt{miy18})-Subaru Strategic Program (SSP) wide-field and deep imaging data~\citep{aih18a}
and archival IR data (ALLWISE).

The magneto-hydrodynamic simulation indicates that a super-Eddington system, which could be realized in DOGs, would produce relativistic jets~\citep{tak10,sad15} 
and also could launch radiatively driven outflows which would eventually produce shock-driven radio emission~\citep[e.g.,][]{zak14}. 
Since previous high Eddington AGN sources were searched mostly by optical~\citep[e.g.,][]{kel13}, an efficient method to search for radio-loud, super-Eddington AGN/DOGs potentially opens up new access to the putative population \citep[e.g.,][]{lon15,pat22}.

Another important aspect for searching such radio-loud DOGs is that radio-loud DOGs can also be
previously missed populations of high-energy cosmic rays (CRs; \citealt{rie22}). The CRs should escape from the jets and wander in the interstellar medium (ISM). These CRs can produce high-energy neutrinos via hadronuclear interactions if the ISM gas density is sufficiently high. Local radio galaxies do not achieve such a high gas density. On the other hand, radio-loud DOGs can provide a gas-rich environment, and thus radio-loud DOGs can be a good candidate for high-energy neutrino sources. 

IceCube Collaboration has been detecting cosmic high-energy neutrinos with an energy range of 30 TeV to a few PeV, the majority of which most likely originate from extragalactic sources~\citep[e.g.,][]{aar13,abb22}. Recently, IceCube Collaboration reported a radio-loud AGN and an AGN with dusty environment as cosmic neutrino source candidates~\citep[e.g.,][]{ice18,Ice22}. 
Radio-loud DOGs exhibit both features, and thus it is worth discussing neutrino emission there.

In this paper, we report a finding of one radio-loud DOG J1406+0102
at $z=0.236$, which shows a super-Eddington accretion signature from both the optical spectral analysis and the wide-range optical-to-IR spectral energy distribution (SED) analysis.

Throughout the paper, we adopt standard cosmological parameters ($H_0 = 70.0$~km~s$^{-1}$~Mpc$^{-1}$, $\Omega_\mathrm{M}=0.3,$ and $\Omega_\Lambda = 0.7$), translating to a scale of 3.75 kpc per arcsec at $z=$ 0.236 of J1406+0102.

\section{Data and Analysis}\label{sec:data}
Radio-loud DOG~J1406+0102 (hereafter J1406) was identified as one of the 29 radio-loud DOGs detected in the VLA/FIRST 1.4~GHz radio survey \citep{hel15} within 1 arcsec positional matching (Fukuchi et al. in preparation), 
and the reader should refer to \cite{nob19} for the original 571 IR-bright DOGs sample. 
After cross-matching all radio-loud DOGs with the SDSS~DR17 catalog \citep{abd22}, only the brightest J1406 ($i_\mathrm{AB} = 18.45$) was left as the source with available optical spectra.
The obtained radio flux density is 9.2~mJy at 1.4~GHz. Taking into account the obtained spectroscopic redshift of $z=0.2367$ (see Section~\ref{sec:spec_fitting}), the flux density corresponds to the radio power of $P_\mathrm{\nu} \sim 1.5 \times 10^{24}$~W~Hz$^{-1}$, which is classified as radio-intermediate or radio-loud AGN \citep[e.g.,][]{gan19}.

\begin{figure*}
    \centering
    \includegraphics[width=0.46\textwidth]{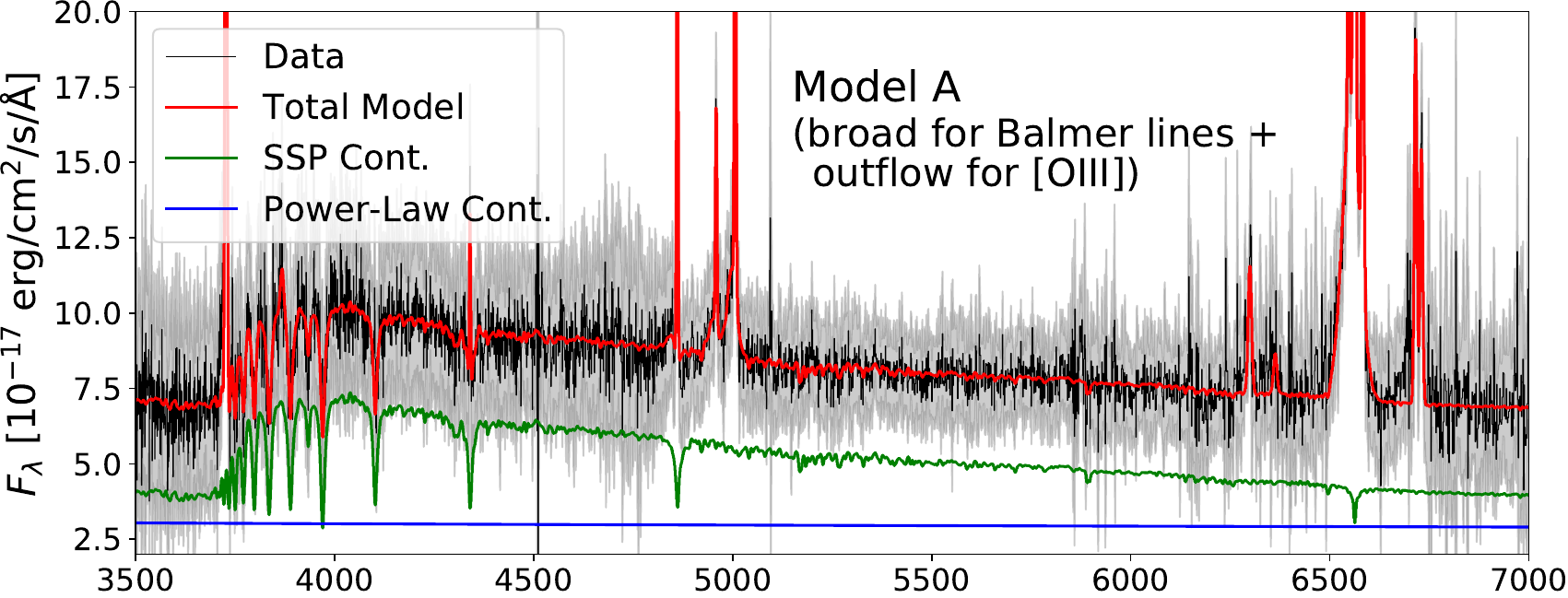}~
    \includegraphics[width=0.26\textwidth]{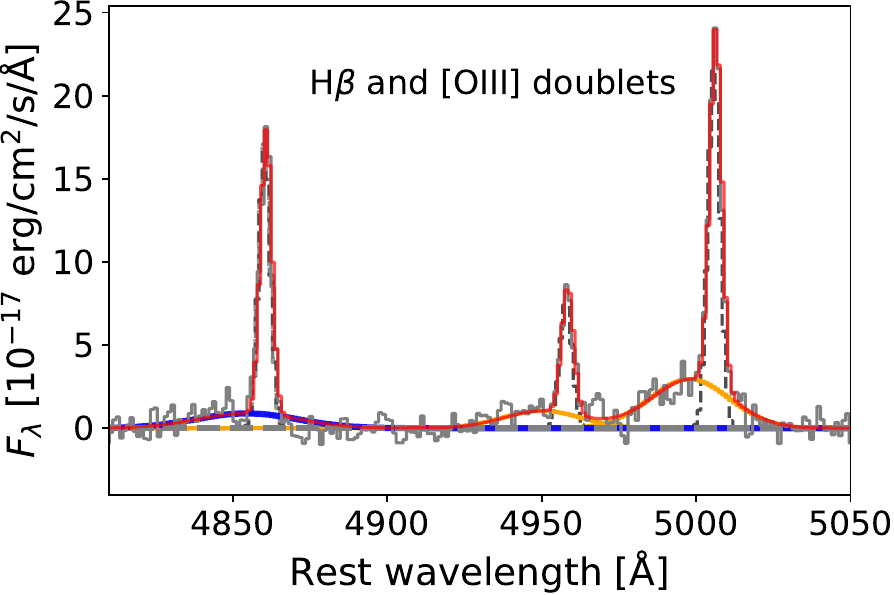}~
    \includegraphics[width=0.26\textwidth]{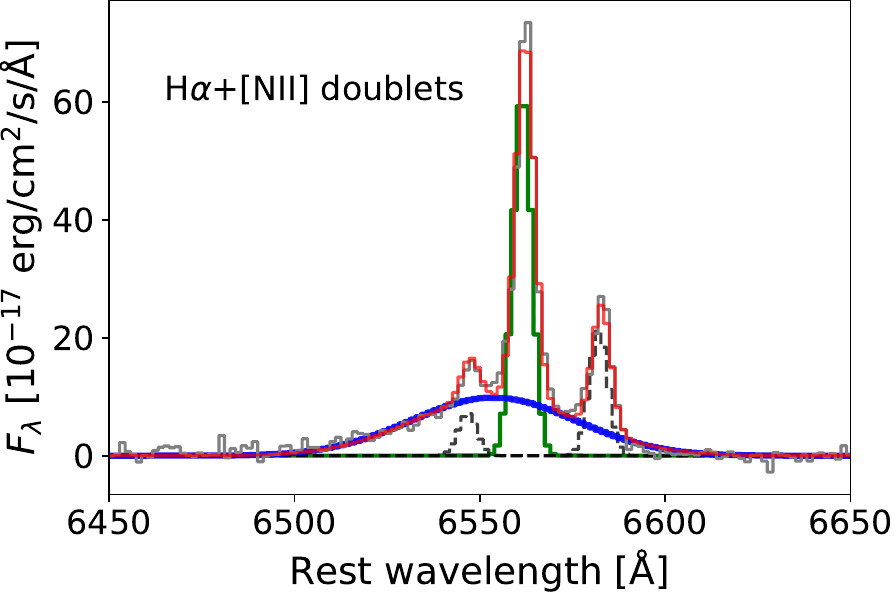}~\\
        \includegraphics[width=0.46\textwidth]{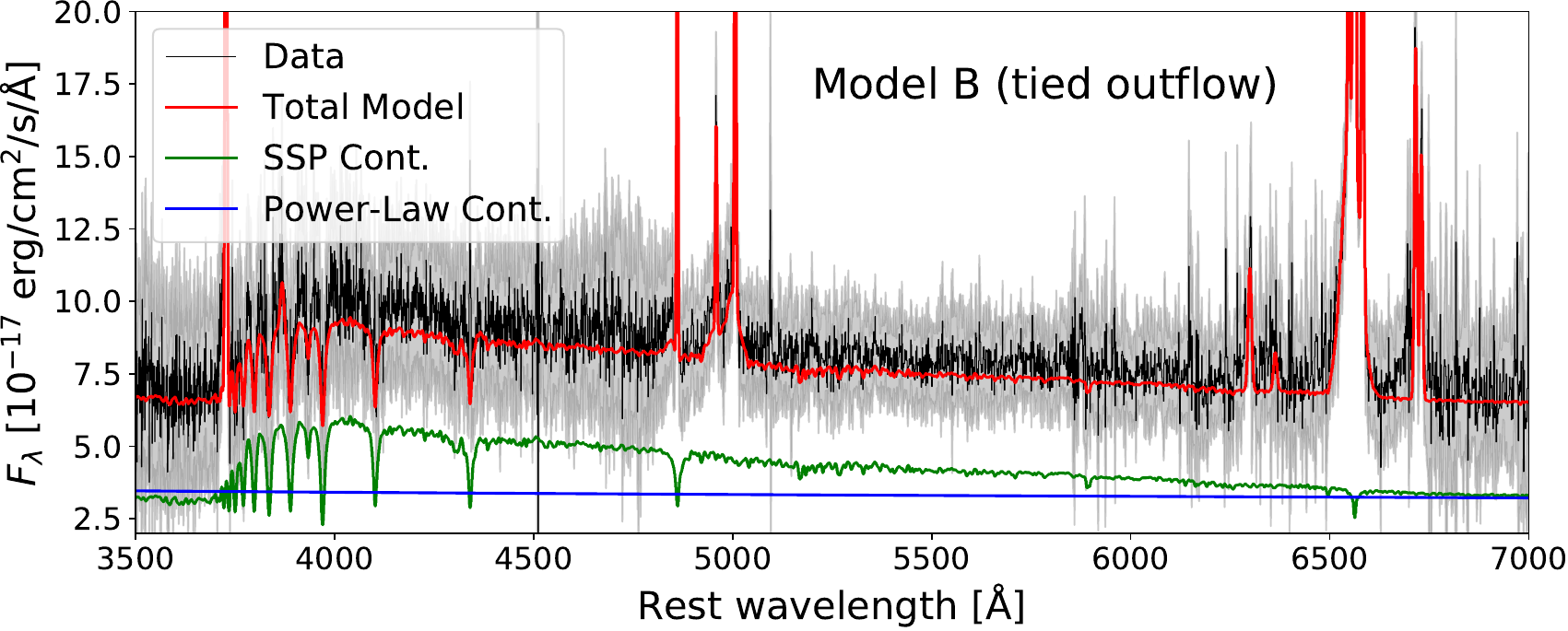}~
    \includegraphics[width=0.26\textwidth]{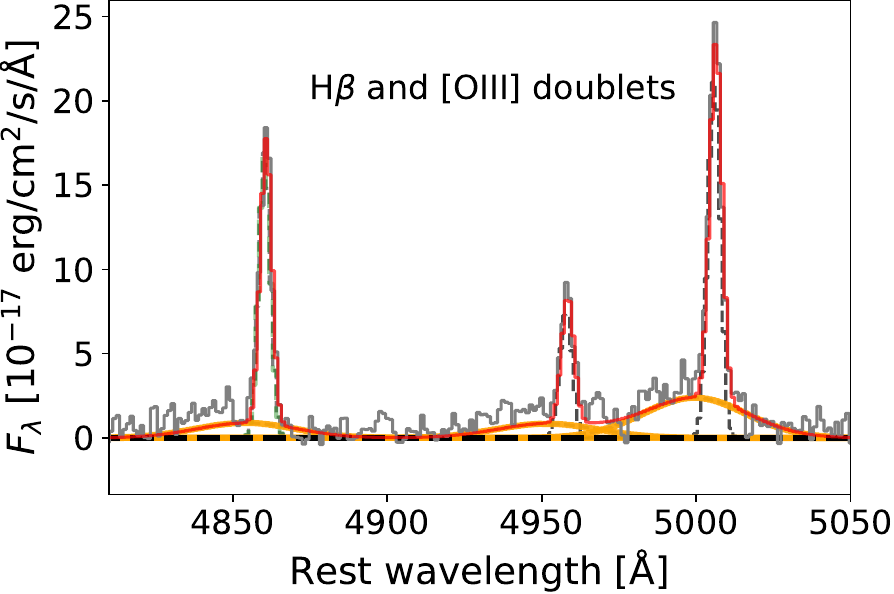}~
    \includegraphics[width=0.26\textwidth]{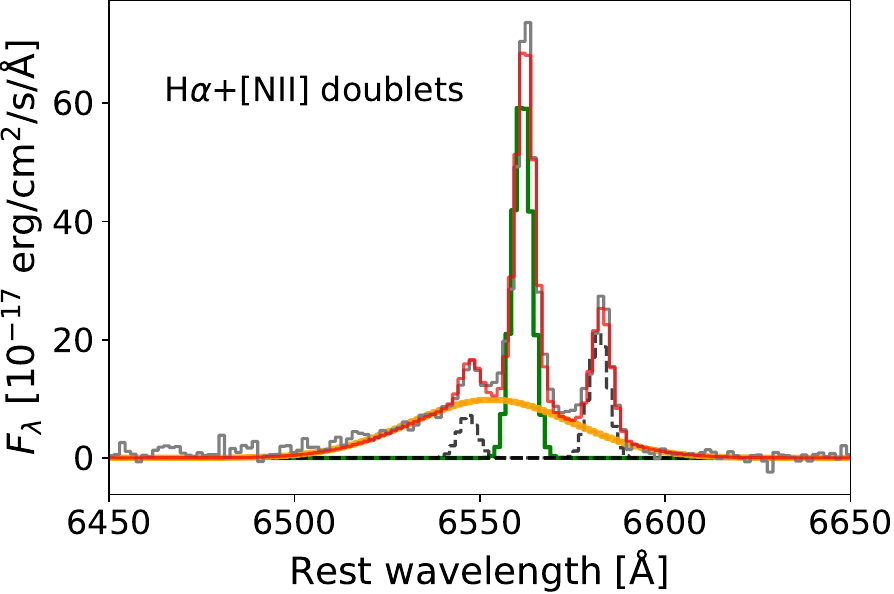}~\\
           \includegraphics[width=0.75\textwidth]{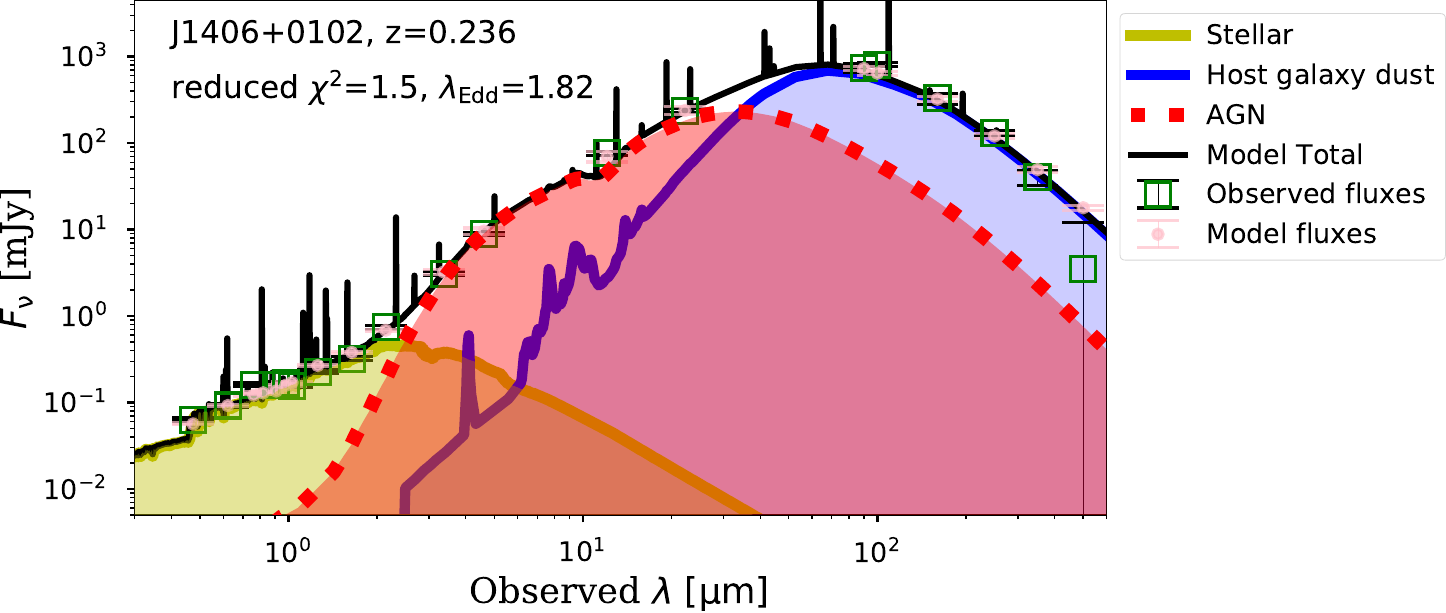}~\\
    \caption{
(Top) Left hand figure shows the SDSS spectrum of J1406 with fitting result using GELATO for model A: untied velocity shift and FWHM for blueshifted [O~{\sc{iii}}] lines and a broad component of each Balmer line. 
SDSS spectrum is shown in black with 3 sigma uncertainty in grey. The simple stellar populations (SSP) continuum and power-law continuum are plotted in green and blue, respectively. The overall fit is shown in red, which include emission line models with SSP continuum and power-law continuum.  
(Top, middle to right) SDSS spectrum fitting results at H$\beta$-[O~{\sc{iii}}] range and H$\alpha$-[N~{\sc{ii}}] range. The spectrum is plotted as a gray line with the narrow components (gray or green dashed line), and the overall fit is shown in red. 
Broad H$\alpha$ and H$\beta$ Gaussian components are plotted in blue, and blueshifted [O~{\sc{iii}}] component is plotted in orange. 
(Middle) Same figure with top panel but with a different model set-up (model B): tied velocity shift and FWHM for [O~{\sc{iii}}] and Balmer lines. 
Right hand figure shows Broad H$\alpha$ and H$\beta$ Gaussian components and blueshifted [O~{\sc{iii}}] component is plotted in orange, which have tied  velocity shift and FWHM. 
(Bottom) SED fitting result of J1406. 
The red dotted line represents the AGN direct and dust emission. The blue solid line represents the dust emission from the host galaxy. 
The yellow line represents the stellar emission from the host galaxy and 
the green dashed line represents the intrinsic stellar emission.
The black solid line is the combined one of dust, stellar, nebular, and AGN.
}\label{fig:spectrum_fitting}
\end{figure*}

Figure~\ref{fig:cutout_image} shows multiwavelength images of J1406, including Subaru/HSC S16a optical 5 bands ($g$, $r$, $i$, $z$ and $y$), VISTA Kilo-Degree Infrared Galaxy Survey (VIKING\footnote{VIKING: \url{https://www.eso.org/sci/observing/phase3/data_releases/viking_dr2.pdf}} \footnote{VIKING release note: \url{http://www.eso.org/rm/api/v1/public/releaseDescriptions/135}})
$Ks$ band, WISE 4 bands (W1, W2, W3, and W4;~\citealt{wri10}), VLA/FIRST and VLA sky survey (VLASS, \citealt{lac20}) images, and their flux densities are summarized in Table~\ref{tab:SED}. 
HSC images show a clear tidal feature thanks to its high angular resolution of $\sim$~0.6 arcsec at $i$-band~\citep{aih18a}, supporting the DOGs formation scenario as a post-phase of galaxy major mergers. Radio images are un-resolved by VLASS, with the spatial resolution of 2.5~arcsec, setting an upper-limit on the physical size of radio emission of $<4.7$~kpc.
We will discuss possible origins of radio emission in Section~\ref{sec:radio_origin}.

\subsection{SDSS Spectrum and Spectral Fitting}\label{sec:spec_fitting}

The top left panel of Figure~\ref{fig:spectrum_fitting} shows the SDSS optical spectrum of J1406 (black solid line). The spectra shows a clear Balmer break at $\sim4000$\AA, suggesting that the continuum is dominated by the stellar direct component from the host galaxy.
On the other hand, the emission lines show ionized lines such as [SII]$\lambda\lambda6716+6730$ doublets, [NII]$\lambda\lambda6583+6548$ doublets and [NeIII]$\lambda\lambda3868.8$, and neutral lines of [OI]$\lambda6300.3+6363.8$ doublets, providing the spectroscopic redshifts of $z=0.2367$.
The middle to right panels exhibit clear broad emission lines
around the Balmer lines and [O~{\sc{iii}}]$\lambda$4959, $\lambda$5007 lines (hereafter [O~{\sc{iii}}] lines), suggesting the existence of AGN.

We performed the spectral fitting using the GELATO code \citep{hvi22} \footnote{Link to the GitHub \url{https://github.com/TheSkyentist/GELATO}}, which builds the composite stellar/AGN spectrum through simple stellar populations (SSP) with an AGN power-law continuum component and narrow/broad emission lines. 
GELATO uses SSP models from the Extended MILES stellar
library~\citep{vaz16} assuming the \cite{cha03} initial mass function(IMF) and isochrones of~\cite{gir00} to produce a continuum of stellar origin for the host galaxy. In addition, GELATO has the ability to check and add a power-law component to account for the continuum component of AGNs: $f_\nu \propto \lambda^\alpha$~\citep[e.g.,][]{van01}.

For emission lines, we added [O~{\sc{i}}]  and [O~{\sc{ii}}] with fixed redshift and velocity dispersions as the origin of star formation. 
We also added [S~{\sc{ii}}], [N~{\sc{ii}}], [O~{\sc{iii}}], [Ne~{\sc{iii}}] and Balmer lines for emission lines originating from AGN or stars. 
Here, according to the default set of \cite{hvi22}, the redshift and velocity dispersion of the Balmer line are not fixed with the other emission lines, while for the other lines the redshift is fixed and the velocity dispersion is variable.
The redshift variation (velocity shift) is~$\pm$~300~km~$\mathrm{s}^{-1}$ and velocity dispersion is from 60~km~$\mathrm{s}^{-1}$ to 500~km~$\mathrm{s}^{-1}$~\citep{hvi22}. 
In addition, for the  [O~{\sc{iii}}] emission complex, an additional outflow component can be tested, which is not forced to share the redshift or dispersion of the narrow components.
Since J1406 shows a clear blue-shifted [O~{\sc{iii}}] line profile, we set the full width at half-maximum (FWHM) up to 2000~km~s$^{-1}$ with the maximum velocity shift of 1000~km~s$^{-1}$ for the outflow component in the [O~{\sc{iii}}] emission lines. 
We also set a maximum velocity shift of 1000~km~s$^{-1}$ for the broad components of the Balmer lines with velocity dispersion range of 500~km~$\mathrm{s}^{-1}$ to 6500~km~$\mathrm{s}^{-1}$. 

We here tested two spectral models, based the differences of the physical origins attributed to the broad emission lines. Model A assumes that broad Balmer lines are virialized one from the broad-line region, while model B assumes that the Balmer lines are associated with AGN outflows. 
For model A (virialized Balmer broad emissions), we used the same FWHM and outflow velocity shift for both H$\alpha$ and H$\beta$, but allowing the [O~{\sc{iii}}] lines to have independent values. This reflects the assumption that the broad Balmer emissions would originate from the broad line region of the AGN, whereas the broad [O~{\sc{iii}}] emission originates from AGN driven outflows. On the other hand, model B (outflow-based Balmer broad emissions) fixes both the velocity shifts and FWHMs for all Balmer lines and [O~{\sc{iii}}], under the assumption that the AGN driven outflows are the dominant source of the broad emission features.

The top panels of Figure~\ref{fig:spectrum_fitting} show the spectral fitting results (for model A) of the SDSS spectra, with the reduced chi-square (reduced $\chi^2$) of 1.2.
The [O~{\sc{iii}}]~emission shows an strong blueshifted outflow feature, with the FWHM of $770\pm 50$ km~s$^{-1}$ and velocity shift of $-550\pm70$~km~s$^{-1}$. 
The broad components of the Balmer lines (H$\beta$ and H$\alpha$) show a large FWHM of $1020 \pm 30$ km~s$^{-1}$ and velocity shift of $-400 \pm 30$~km~s$^{-1}$.
The observed luminosities of the broad H$\alpha$ and H$\beta$ are $\log \left(L_{\mathrm{H}\alpha}/\mathrm{erg}~\mathrm{s}^{-1}\right) = 41.99\pm0.01$ and $\log (L_{\mathrm{H}\beta}/\mathrm{erg}~\mathrm{s}^{-1}) = 40.81\pm0.01$, respectively.
The extinction correction, $E(\mathrm{B−V}) = 1.35$, is obtained using the Balmer decrement method, applying the \citet{cal00} extinction law and assuming an intrinsic H$\alpha$/H$\beta$ ratio of 3.06 \citep{don08} for the broad components.

Assuming that the Balmer broad components are virialized, we calculate the H$\alpha$ and H$\beta$ based virial BH masses using the following relations \citep{gre05}:
\begin{align}
M_{\mathrm{BH}} &= (2.0^{+0.4}_{-0.3}) \times 10^6 \left( \frac{L_{\mathrm{H}\alpha}}{10^{42}~\mathrm{erg~s^{-1}}} \right)^{0.55 \pm 0.02} \nonumber  \\
&\quad \times \left( \frac{\mathrm{FWHM}_{\mathrm{H}\alpha}}{10^3~\mathrm{km~s^{-1}}} \right)^{2.06 \pm 0.06} M_{\odot}.
\end{align}
and
\begin{align}
M_{\mathrm{BH}} &= (3.6 \pm 0.2) \times 10^6 
\left( \frac{L_{\mathrm{H}\beta}}{10^{42}~\mathrm{erg~s^{-1}}} \right)^{0.56 \pm 0.02} \nonumber \\
&\quad \times \left( \frac{\mathrm{FWHM}_{\mathrm{H}\beta}}{10^3~\mathrm{km~s^{-1}}} \right)^2 M_{\odot}.
\end{align}
The resulting BH mass is $\log\ (\mbh/\msun)=7.30 \pm 0.25$ ($\log\ (\mbh/\msun)=7.40 \pm 0.25$) for H$\alpha$ (H$\beta$), respectively. Here, the 0.25~dex uncertainty includes the 0.2 dex intrinsic scatter of the scaling relation and the uncertainty of spectral fitting. 
Since both values are consistent within the scatter, we apply H$\alpha$ based BH mass as a fiducial value in this study\footnote{We note that caution is needed when estimating the BH mass based on the virial BH mass method due to the inherent uncertainties involved. According to \cite{Ber24}, the BH mass derived by the virial method could achieve an error of about 1 dex in some cases. In any case, we also note that the Eddington ratio of J1406 would be high enough around the Eddington limit, so this uncertainty does not significantly change our conclusions. Additional observations, such as measuring the dynamical mass, would reduce this uncertainty \citep{abu24}.}.

The AGN bolometric luminosity ($L_\mathrm{bol}$), estimated from the extinction-corrected [O~{\sc iii}] luminosity ($L_\mathrm{[OIII],abs}$), is $\log (L_\mathrm{bol,[OIII]}/\mathrm{erg}\mathrm{s}^{-1}) = 45.91\pm0.38$. This value is consistent with $L_\mathrm{bol}$ derived from the SED fitting (Section~\ref{sec:SED_fit}). Here, we adopt the relation $L_\mathrm{bol,[OIII]} = 3500 L_\mathrm{[OIII],abs}$ with a scatter of 0.38~dex \citep{hec04}, which dominates the uncertainty in $L_\mathrm{bol,[OIII]}$. The estimated Eddington ratio is $\lambda_\mathrm{Edd} = 3.3^{+6.1}_{-2.1}$, indicating a super-Eddington accretion regime.

In contrast, the broad components of the Balmer lines may not be virialized and could instead be associated with an outflow launched from the accretion disk. Model B explores this possibility, by freezing its velocity shift and FWHM to those of the [O~{\sc iii}] outflow component (see also the second top panels of Figure~\ref{fig:spectrum_fitting}). The resulting reduced $\chi^2$ of 1.25 is comparable to that obtained when fitting with a broad Balmer component.

Under this scenario, J1406 would be classified as a type-2 AGN exhibiting strong ionized outflows (FWHM = $1000 \pm 30$~kms$^{-1}$, $v_\mathrm{shift} = 420 \pm 20$~kms$^{-1}$). Consequently, BH mass estimates based on the virial method may not be reliable for J1406, making a stellar mass-based BH mass estimation more crucial (see Section~\ref{sec:SED_fit}).

\begin{deluxetable}{cccccccc}
\tabletypesize{\footnotesize}
\tablecolumns{5}
\tablewidth{0pt}
\tablecaption{Optical to FIR flux densities for J1406+0102}\label{tab:flux1}
  \tablehead{
      \colhead{Catalog} & 
      \colhead{Band} &  
      \colhead{$\lambda_{\mathrm{center}}$} & 
      \colhead{Flux Density} &
      \colhead{Ref.} 
      \\
       \colhead{} & 
       \colhead{} & 
       \colhead{($\mathrm{\mu m}$)} &
       \colhead{(mJy)}&
       \colhead{}
     }
\startdata
   Subaru/HSC &       $g$ &         0.5 &   0.06 +/- 0.003 &  [1], [2] \\
   SSP S19A       &       $r$ &         0.6 &   0.09 +/- 0.005 &           \\
            &       $i$ &         0.8 &   0.16 +/- 0.008 &           \\
            &       $z$ &         0.9 &   0.16 +/- 0.008 &           \\
            &       $y$ &         1.0 &   0.17 +/- 0.008 &           \\
                 \hline 
 VIKING DR2 &       $Z$ &         0.8 &  - &  [3] \\
            &       $Y$ &         1.0 &    0.16 +/- 0.01 &           \\
            &       $J$ &         1.3 &    0.23 +/- 0.01 &           \\
            &       $H$ &         1.6 &    0.32 +/- 0.02 &           \\
            &      $K_\mathrm{s}$ &         2.1 &    0.74 +/- 0.04 &           \\
                 \hline 
    ALLWISE &      W1 &         3.4 &      3.1 +/- 0.2 &       [4] \\
            &      W2 &         4.6 &      8.9 +/- 0.4 &           \\
            &      W3 &        12.1 &     75.8 +/- 3.8 &           \\
            &      W4 &        22.2 &   236.2 +/- 11.8 &           \\
                 \hline 
     HATLAS &     100 &         100 &   801.4 +/- 41.3 &      [5], [6] \\
            &     160 &         160 &   328.7 +/- 47.9 &           \\
            &     250 &         250 &    130.6 +/- 7.3 &           \\
            &     350 &         350 &     40.5 +/- 8.0 &           \\
            &     500 &         500 &      3.5 +/- 8.6 &           \\
                 \hline 
      AKARI &  Wide-S &        90.0 &   737.0 +/- 86.0 &      [7] \\
\enddata
\tablecomments{
The optical to FIR data for J1406+0102 used in 
the SED fitting. Fluxes and their central wavelengths $\lambda_{\mathrm{center}}$ are obtained from references.
References: [1] \cite{aih18a}, [2] \cite{aih18b};
[3] \cite{edg13};
[4] \cite{wri10}; 
[5] \cite{pil10}, [6] \cite{smi17};
[7] \cite{ona07}.
}
\label{tab:SED}
\end{deluxetable}


\begin{deluxetable}{lcc}
\tablecaption{Parameter sets used in \texttt{CIGALE} for the SED fitting of J1406}
\tablewidth{0.48\textwidth}
\tablehead{
\colhead{Model-Palameter} 
 & \colhead{Values} & \colhead{Select$^a$}
}
\startdata
   \multicolumn{2}{c}{SFH: delayed SFH with a $\tau$ decay burst}\\
  $\tau$ of the main population (Myr)& 1000, 3000, 5000 & 5000\\
    Age of the main population (Myr)& 1000, 3000, 5000& 1000\\
    & 10,000 &\\
      $\tau$ of the starburst population (Myr)& 30, 100 & 100\\
    Age of the late burst (Myr)& 10, 30, 100 & 10\\
    Mass fraction of burst population & 0.001,0.01,0.1,0.3 & 0.3\\
     \hline 
     \multicolumn{2}{c}{Stellar population synthesis model: (1) and (2)}\\
    Metallicity (Z) & 0.4 $Z_{\odot}$, $Z_{\odot}$ & $Z_{\odot}$\\
     \hline 
    \multicolumn{2}{c}{Dust attenuation: (3).}\\
   E(B−V) for young stars continuum& 0.2-2.0 (step 0.3) & 1.1\\
   E(B−V) old factor &  0.1, 0.3, 0.9 & 0.1\\
    \hline 
    \multicolumn{2}{c}{Dust emission: (4)} \\
    $\alpha$ slope in $dM_\mathrm{dust} \propto U^{-\alpha} dU_\mathrm{intensity}$ &  0.75 &  \\
    (IR power-law slope)\\
     \hline 
    \multicolumn{2}{c}{AGN (UV-to-IR): SKIRTOR (5).}\\
 Viewing angle ($\theta$)  & $30^{\circ}$, $70^{\circ}$ & $70^{\circ}$ \\
AGN fraction in total IR luminosity & 0.01, 0.3, 0.5,  & 0.5\\
& 0.7, 0.9 & \\
 Extinction law of polar dust & SMC &\\
 E(B−V) of polar dust & 0.03, 0.1, 0.15 & 0.03\\
Temperature of polar dust (K) & 100, 200, 300 & 200\\
     \hline 
\enddata
  \thispagestyle{empty}
  \tablenotetext{a}{The finally selected parameter value by the SED fitting.}
\tablenotetext{}{
References: (1) \cite{bru03}; (2) \cite{cha03}; (3) \cite{cal00}; (4) \cite{dal14}; (5) \cite{sta16}.
} 
\label{tab:para}
\end{deluxetable}


\subsection{Optical--FIR SED Fitting}\label{sec:SED_fit}
We apply the SED decomposition method to obtain the AGN and host galaxy components
by using the \texttt{CIGALE} SED fitting code \citep[e.g.,][]{yan22}, which builds the composite stellar/AGN spectrum through simple stellar populations (SSP) with flexible star-formation history (SFH) and AGN radiation.

We first construct the multi-wavelength SED of J1406,
covering from optical to 500~$\mu$m, by using the dataset from the Subaru HSC SSP S19A data ($g$, $r$, $i$, $z$ and $y$-bands), VIKING $J$, $H$, and $Ks$-bands, WISE IR data (3.4, 4.6, 12, and 22 $\um$), AKARI FIS 90~$\um$ \citep{kaw07}, and Herschel Astrophysical Terahertz Large Area Survey (H-ATLAS; \citealt{pil10}) 100, 160, 250, 350, and 500 $\mu$m. 
Here, for the HSC data points, we use the cModel magnitudes rather than the PSF magnitudes to capture the total emission from both the AGN and host galaxies \citep[see detailed discussions on the cModel magnitude;][]{aih18a}. This is because
other datasets (e.g., WISE and AKARI) do not provide spatially resolved components and
thus include dust emission from both the AGN and host galaxies.
For the WISE data points, we use the profile-fitting magnitudes~\citep{wri10}.
We have also applied Galactic extinction correction~\citep{sch98} to the obtained photometries in the HSC and VIKING bands.
Their flux densities and references are summarized in Table~\ref{tab:SED}.
As summarized in Table 1, the optical to near-IR color slope is $\alpha = 1.55 \pm 0.13$, defined by $f_\nu \propto \lambda^\alpha$. This value is redder than the typical quasar value of $\alpha=0.5$ in this wavelength range \citep[e.g.,][]{van01}, suggesting that the SED is dominated by either the host galaxy or dust-obscured AGN.

We then utilize the latest version of the \texttt{CIGALE} SED fitting code called the \texttt{CIGALE} 2022.0 \citep[hereafter \texttt{CIGALE};][]{yan22}. 
We refer the reader to \cite{yan20,yan22} for a full description of \texttt{CIGALE}. 
The models and parameter grids used in our analysis are summarized in Table \ref{tab:para}, with all other parameters set to their default \texttt{CIGALE} values.  
The grids of these models are fitted to the observational data in \texttt{CIGALE}, and \texttt{CIGALE} estimates the reduced chi-square for each parameter.
Physical values are based on the likelihood-weighted means of the parameters obtained by the \texttt{CIGALE} fittings, and errors are based on the standard derivations of the parameters obtained.
Here, we describe the main steps of model building for fitting the SED of J1406 covering the optical band to the FIR band and the results of the fitting.

\subsubsection{Host Galaxy Component}\label{sec:sfh}
The host galaxy component can be characterized by the combination of the assumed SFH and the initial mass function (IMF) of the stellar emission.
On the SFH, we applied the delayed SFH with optional exponential burst to characterize the SED of ULIRGs with the experience of recent starbursts \citep[e.g.,][]{rie09,cie15, boq19}.

The stellar emission is modeled using the \cite{bru03}, and we adopt the \cite{cha03} IMF. 
The metallicity of the gas is set to 0.4~$Z_{\odot}$ and $Z_{\odot}$ to account for the possibility that J1406 is in a metal contamination phase and that the metal fraction is smaller than the metallicity of the Sun ($Z_{\odot}$).

The stellar emission was attenuated by the \cite{cal00} attenuation law. 
The parameter range for dust attenuation is set wider range, E(B-V)~$= 0.2-2.0$ with a step of 0.3 to match the current redder SED colors (see Section~\ref{sec:SED_fit} and Table~\ref{tab:SED}). 
The IR SED of the dust heated by stars was implemented with the \cite{dal14} template.
We added nebular-emission components to the SED using the ‘nebular-emission’ module.

\subsubsection{AGN Component}\label{sec:agn}

For the AGN accretion disk and dust component,
\texttt{CIGALE} utilizes SKIRTOR \citep{sta12,sta16} AGN model, which is one of the clumpy two-phase torus models based on 3D radiation-transfer code
and this model covers from the UV-to-far-IR emission of the observed AGN \citep{ich12,ich15,ich19,ric23}. 
The SKIRTOR AGN model covers the most recently updated polar AGN dust emission that has recently been suggested from the MIR interferometric observations \citep[e.g.,][]{bur13,hon12,hon13,hon19} as well as recently reported JWST-discovered AGN \citep[e.g.,][]{li25}.
Although the dust grain size distribution is not yet known for the polar dust emission \citep[but see][for the detailed discussion of the dust size distribution]{lyu18,taz20a,taz20b},
here we adopted the Small Magellanic Cloud (SMC) extinction curve \citep{pre84} for the polar dust extinction curve, since it is preferred from AGN observations \citep[e.g.,][]{hop04,sal09,bon12} and it is also the default manner of \texttt{CIGALE}.
We set the polar dust temperature at 100, 200, 300~K for tracing the MIR AGN dust emission peak with an emissivity index of 1.6 \citep{yan20}.

\subsubsection{SED fitting result}\label{sec:results}
The bottom panel of Figure~\ref{fig:spectrum_fitting} shows the SED fitting result for the best model of J1406 by \texttt{CIGALE}, with the decomposed AGN (red dotted line), stellar (yellow solid line), and host dust (blue solid line) components. 
The SED fitting provides the total stellar mass of $\log (\mstar/\msun)= 9.94 \pm 0.02$.
Using the local scaling relation between $\mbh$ and the bulge mass~\citep{kor13} with the redshift evolution of $\mbh/\mstar \propto (1+z)^{0.68}$ \citep{mer10}, and assuming the total stellar mass as the bulge mass for J1406, we estimate a BH mass of 
$\log\ (\mbh/\msun) = 7.51 \pm 0.28$.
This BH mass estimation
is also roughly consistent with the H$\alpha$-based value of $\log\ (\mbh/\msun) = 7.30 \pm 0.25$ (see Section~\ref{sec:spec_fitting}). 
The \texttt{CIGALE} fitting also gives the 
AGN bolometric luminosity of $\log\ (L_\mathrm{AGN,bol}/\mathrm{erg}~\mathrm{s}^{-1}) =  45.87 \pm 0.02$, resulting in an Eddington ratio of 
$\lambdaedd = 1.82^{+1.81}_{-0.91} \gtrsim 1$. 
Based on the results above, our main conclusion is that J1406 is in a super-Eddington accretion phase by using both SDSS spectral fitting (Section~\ref{sec:spec_fitting}) and SED fitting methods (Section~\ref{sec:SED_fit}). 
This result also supports the idea that the radio-loud DOGs search would be a good pathway to find rapidly growing SMBHs in a dust-obscured phase. This is further discussed in the following paper (Fukuchi et al. in preparation), and \cite{nob22} also reported some super-Eddington BHs among our DOGs sample. 

\section{Discussion}\label{sec:dis}

\begin{deluxetable*}{cccccccc}
\tabletypesize{\footnotesize}
\tablecolumns{5}
\tablewidth{0pt}
\tablecaption{Summary of J1406+0102 radio flux densities along with information from radio surveys\label{tab:flux2}}
  \tablehead{
      \colhead{Catalog} & 
      \colhead{$\nu_{\mathrm{center}}$} & 
      \colhead{$\Delta_\nu$} & 
       \colhead{PSF} &
      \colhead{$\sigma_\mathrm{rms}$} &
      \colhead{Flux Density} &
            \colhead{Date} &
      \colhead{Ref.} 
      \\
       \colhead{} & 
       \colhead{(GHz)} &
       \colhead{(GHz)} &
       \colhead{(arcsec)} &
      \colhead{mJy beam$^{-1}$} & 
       \colhead{(mJy)}&
                   \colhead{} &
       \colhead{}
     }
\startdata
 TGSS &         0.155 & 0.016 & 25 & 3.5 &  $<$ 24.5 & 2010 April $\sim$ 2012 March &  [1] \\
                 \hline               
   FIRST &         1.4 &  0.03 & 5 & 0.15 & 9.19 $\pm$ 0.15 & 1998 July &  [2] \\
                 \hline 
 VLASS(epoch1)&              3.0 &  2 & 2.5 & 0.15 & 7.89 $\pm$ 0.28 & 2019 April & [3],[4] \\
   VLASS(epoch2)    & & &      &           &  7.92 $\pm$ 0.29 & 2021 November &  \\ 
\enddata
\tablecomments{We show radio flux information of J1406+0102. 
 J1406 is not included in the TGSS ADR1 catalog with a detection threshold of 7$\sigma$.
References: [1] \cite{int17}; [2] \cite{bec95}; [3] \cite{lac20}; [4] \cite{gor21} 
}\label{tab:radio_spectral_index}
\end{deluxetable*}

\begin{figure}
    \centering
    \includegraphics[width=0.48\textwidth]{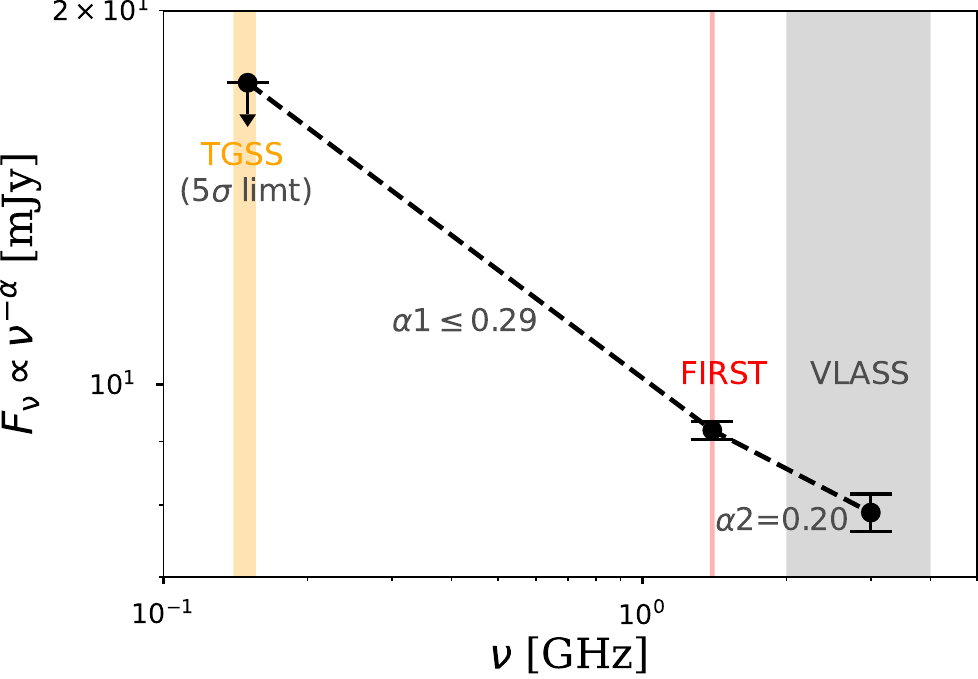}~
    \caption{Radio SED of J1406+0102. $\alpha_1$ ($\alpha_2$) represents the spectral index calculated between the TGSS and FIRST data (and between FIRST and VLASS), respectively. The shaded region corresponds to the bandwidth of each survey, as summarized in Table~\ref{tab:flux2}.}\label{fig:radio_spectral_index}
\end{figure}

\subsection{Origin of Radio Emission}\label{sec:radio_origin}

Our results strongly indicate that J1406 has a central engine that reaches the super-Eddington accretion. In addition to the bright optical-IR emission, which aligns with the selection criteria for radio-loud DOGs,
J1406 also shows prominent radio emissions observed in both the VLA/FIRST (1.4~GHz) and the VLASS survey $\sim3$~GHz (S-band: 2–4 GHz: \citealt{lac20}), reaching 
$L_\mathrm{1.4GHz} = (2.03 \pm 0.03) \times 10^{40}$~erg~s$^{-1}$.
This is worth to note, since J1406 displays an unusually bright radio feature not commonly seen in previously reported AGN in super-Eddington
\citep{gru04,kom08,liu21,tor23}, except several cases \citep{kom18,ich21,ich23}.
Given the classification as a radio-loud DOG, we here explore the possible origins of these radio emissions: 1) star formation, 2) radio jet, and 3) outflow-driven shock.

\subsubsection{Star Formation}
Host galaxies are not radio silent. They produce both synchrotron radiation from supernova remnants and free-free emission from HII regions, whose luminosities depend on the host galaxy SFR.
The star formation rate for J1406 is estimated from the SED fitting with $\log ( \mathrm{SFR}/\msun~\mathrm{yr}^{-1}) = 2.31 \pm 0.02$. The expected radio emission is  $L_\mathrm{1.4 GHz, SF}=(1.81 \times 10^{37} \times (\mathrm{SFR}/M_\odot ~\mathrm{yr}^{-1})$~erg~s$^{-1} = 3.7\times 10^{39}$~erg~s$^{-1}$~\citep{bel03},
which could contribute $\sim20$\% of the total radio emission at 1.4~GHz,
which is not enough to account for the total radio flux.
This indicates that a significant fraction of the radio emission
should originate either from the AGN radio jet and/or outflow-driven shock.

\subsubsection{Radio Jet} 
Theoretical simulations predict that a super-Eddington accreting BH would produce a radiation-driven jet \citep[e.g.,][]{ohs09,tak10,sad15}. J1406 shows a compact VLASS image with a resolution of 2.5~arcsec,
corresponding to a radius of $<4.7$ kpc at 3~GHz (Figure~\ref{fig:cutout_image}). 
If J1406 hosts a jet, by assuming the jet size of 4.7 kpc, an angle of 45$^\circ$ to the line of sight, and a typical expansion speed of radio lobes~\citep[$\sim0.2c$;][]{nag06,ich16,ich19b}, the estimated kinematic age of the jet
is $\sim 1.5 \times 10^5$~yr.
This relatively young jet (compared to FRI/II sources of $\sim10^8$~yr) is consistent with a jet launched during the current super-Eddington phase, which is expected to last only the order of Myr \citep[e.g.,][]{kaw04,eil20,shi20,ina22,ina25}.
This young jet age is also consistent with the properties of
giga-hertz peaked spectrum (GPS) radio galaxies, which host young AGN jets with linear sizes of 0.1--1 kpc, and radio emission in highly obscured gas environments~\citep[e.g.,][]{ode98}. 
Similar observational trends have also been reported in the radio spectra of heavily dust obscured radio quasars at $z=1$--$2$ \citep[e.g.,][]{lon15,pat20,pat22}.
We also obtain the multi-epoch VLASS data, where the two VLASS 3~GHz observations taken over a span of ~$2.5$~yr show consistent flux densities (see Table~\ref{tab:radio_spectral_index}).
This further suggests that the dominant radio emission is not from the accretion disk, but more likely a radio jet with a size of $\gtrsim1$~pc.

Figure~\ref{fig:radio_spectral_index} shows that the radio spectral index of J1406 is relatively flat, with a slope of $\alpha=-0.2$ for $S_{\nu} \propto  \nu^{\alpha}$ between 1.4~GHz and 3~GHz. This slope is flatter than that would be expected for extended optically thin emission~\citep{ode98}, suggesting synchrotron self absorption (SSA) at $\sim$~GHz range. This is consistent with the highly obscured gas environment of DOGs.
In addition, the Tata Institute of Fundamental Research Giant Meterwave
Radio Telescope Sky Survey~\citep[TGSS;][]{int17} does not detect J1406 at the 7$\sigma$ level ($=24.5$~mJy) at 150 MHz. 
This non-detection is consistent with the expected emission of 14~mJy at 150~MHz, assuming a slope of $\alpha=-0.2$.
Although the current dataset does not have sufficient depth to determine the radio slope over the $\sim100$~MHz to $3$~GHz bands, 
the on-going LoFAR Two-metre Sky Survey (LoTSS) will observe the entire Northern field \citep[$\mathrm{Dec}>0^\circ$;][]{shi22}, which includes the location of J1406 ($\mathrm{Dec}=1.048^\circ$). Once the data become publicly available, the LoTSS survey will achieve a detection limit down to $f_\nu \simeq 0.1$~mJy at 140~MHz, which is expected to be deep enough to detect J1406.

The radio luminosity of J1406 corresponds to the jet power of $L_\mathrm{jet}\sim 10^{44}$~erg~s$^{-1}$~\citep{cav10}, which is $\sim$~1\% of the radiation energy, $L_\mathrm{jet}/L_\mathrm{AGN,bol}\sim 0.01$.  
This jet efficiency is slightly lower than the typical SDSS radio-loud quasars of 3-40\%~\citep[e.g.,][]{ino17} and the
case for rapid spinning BH of 100\%~\citep{tch11}, indicating that J1406 may still have a small BH spin with the BH mass and spin assembly phase.

\subsubsection{Outflow-driven Radio Shock}
Some studies suggest that a rapidly accreting system around SMBHs could launch radiatively driven outflows that propagate into the interstellar medium with shock, and it would eventually produce shock-driven radio emission due to synchrotron radiation~\citep[e.g.,][]{zak14}.

\begin{figure}
    \centering
    \includegraphics[width=0.48\textwidth]{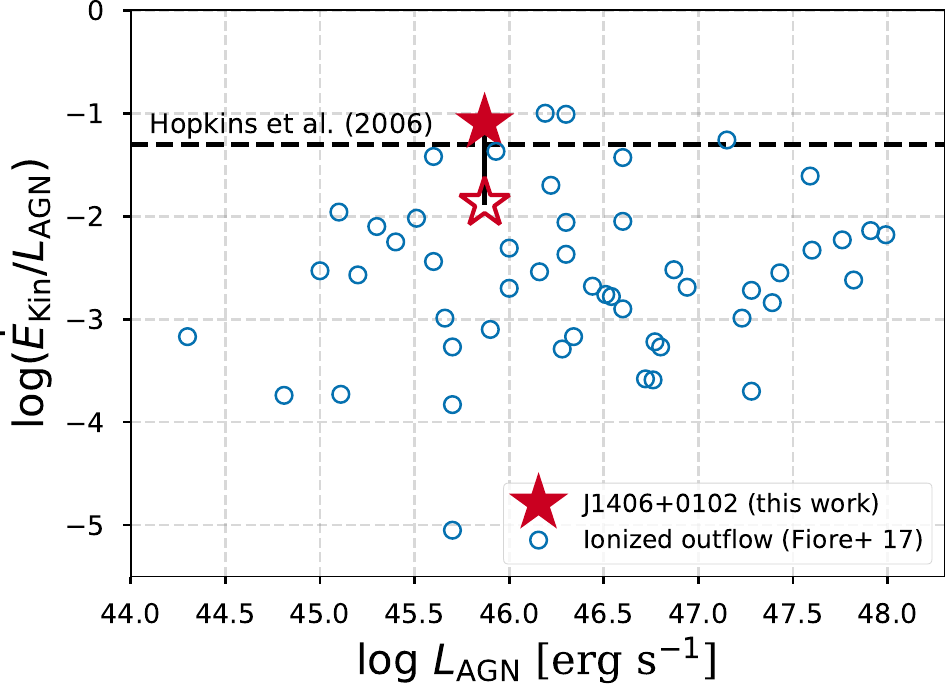}~
    \caption{Efficiency of kinetic luminosity, $\eta_\mathrm{kin}=E_\mathrm{kin}/(L_\mathrm{bol,AGN}\tau)$, as a function of bolometric AGN luminosity for J1406 together with the AGN sample with wide luminosity range compiled by \cite{fio17}. 
    The lowest $\eta_\mathrm{kin}$ of J1406 is estimated using ionized gas radius of $R=$~6~kpc.  
    Cosmological simulation require $\sim$~5\% of radiation energy couples with ISM (black dashed line) to reproduce observed non-stat forming massive galaxies, and J1406 can exceed/comparable to this value.
}\label{fig:kinetic_power}
\end{figure}

Utilizing the [O~{\sc{iii}}] ionized gas outflow, one can estimate the contribution of outflow-driven radio emission \citep[e.g.,][]{zak14}. 
\cite{hwa18} showed a close correlation between radio luminosity and the [O~{\sc{iii}}] outflow velocity of
 $\log (L_\mathrm{1.4GHz,outflow}/\mathrm{erg~s^{-1}})=2\times \log(w_{90}/\mathrm{km~s^{-1}})+33.80$, where $w_{90}$ is the 90\% enclosed velocity and $w_{90} \sim 1.3\times \mathrm{FWHM}$ for the Gaussian distribution. 
 In the case of J1406, $w_{90}= 1080$~km~s$^{-1}$ and 
the estimated 1.4~GHz radio luminosity reaches 
$\log (L_\mathrm{1.4GHz,outflow}/\mathrm{erg~s^{-1}}) = 39.87 \pm 0.56$, which could describe the observed 1.4~GHz radio luminosity for J1406 of
$\log (L_\mathrm{1.4GHz}/\mathrm{erg~s^{-1}}) = 40.30$ within a scatter.
Thus, most of the radio emissions would originate from the AGN activity either from the jet and/or outflows, and in either case, bright radio emissions for DOGs can be a good indicator of rapidly growing SMBHs that reach super-Eddington accretion. 

\subsection{AGN outflow and feedback}

Our results on the optical spectral fitting show that J1406 has a strong ionized [O~{\sc{iii}}] outflow.
This ionized gas velocity well exceeds the 
escaping velocity of the galaxy halo with 
the stellar mass of J1406 of $\log (\mstar/\msun)= 9.94 \pm 0.02$ \citep[e.g.,][]{mos13,che20}.
This indicates that the super-Eddington phase of J1406 (and possibly the radio-loud DOGs in general) has a strong impact on the host galaxy growth by injecting a large amount of energy from its nucleus. 
Indeed, the estimated kinetic energy is $E_\mathrm{kin}=M_\mathrm{HII} v_\mathrm{outflow}^2/2=2.0 \times 10^{58}$~erg with efficient conversion of the radiation energy ($\eta_\mathrm{kin}$) as described below. 
Here, $M_\mathrm{HII}$ is the mass of ionized hydrogen obtained from the outflow components of [O~{\sc{iii}}] with [O~{\sc{iii}}]/H$\beta$ ratio of 1.3 and an electron density of $n_\mathrm{e}=200~\mathrm{cm}^{-3}$ following the estimation by \cite{fio17}. 
Using the typical outflow time scale $\tau=R/v_\mathrm{outflow}$, we obtain a high efficiency of kinetic luminosity of $\eta_\mathrm{kin}=E_\mathrm{kin}/\tau L_\mathrm{AGN,bol} = 8$\% (Figure~\ref{fig:kinetic_power}),
where $R$ is the ionized gas region from the center of the BH, and we assume typical value of~1~kpc~as a fiducial value~\citep[e.g.,][]{zak16}. 
In addition to this value, $R$ can have a maximum value of 6~kpc based on maximum radius of the AGN dominated region for [O~{\sc{iii}}] emission line~\citep{sun18}. 
In this case, $\eta_\mathrm{kin}$ can be $ \sim 1\%$~(open star in Figure~\ref{fig:kinetic_power}).

Figure~\ref{fig:kinetic_power} indicates that $\eta_\mathrm{kin}$=1--8\% for J1406 is located at the upper end of the previously reported values among AGN and quasars~\citep[e.g.,][blue open circles]{fio17}, whose uncertainty is also not small, up to around~1~dex.
This conversion efficiency is sufficient to quench star formation in massive galaxies and also fulfills the requirement in cosmological simulation to reproduce the observed BH-host galaxy scaling relation \citep[5\% of radiation energy couples with ISM: e.g.,][black dashed line in Figure~\ref{fig:kinetic_power}]{hop06}. 
Thus, our result indicates that the existence of the super-Eddington accretion phase can strongly affect the future galaxy stellar mass assembly because of its efficient feedback.

\begin{figure}
    \centering
    \includegraphics[width=0.48\textwidth]{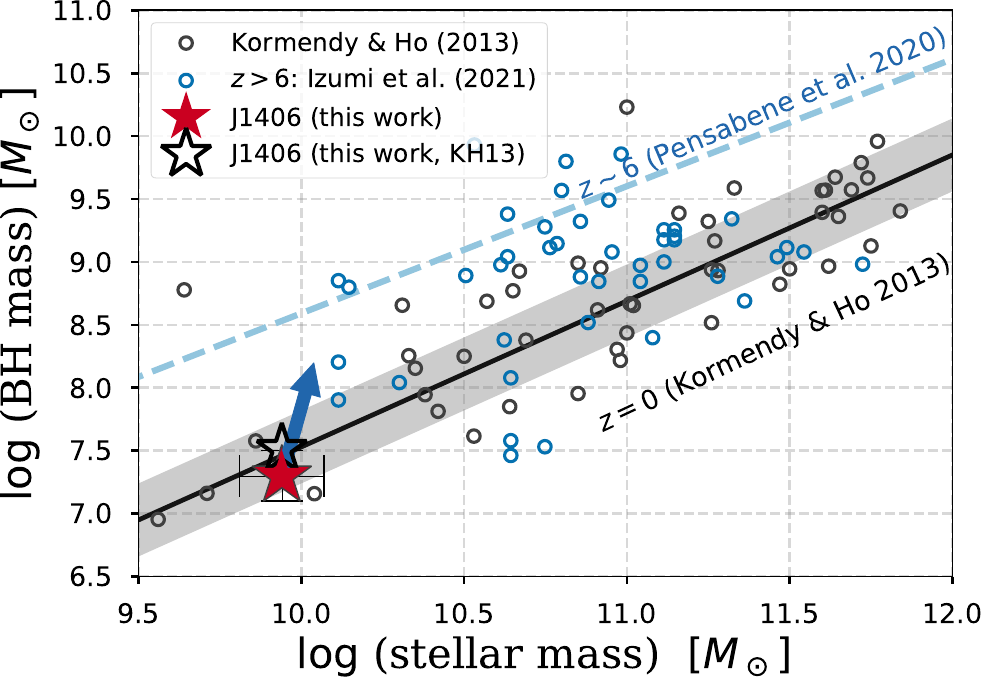}~
    \caption{
    Expected future growth of J1406 in the plane of the BH mass and host stellar mass, together with the distribution of $z > 6$ quasars compiled in~\cite{izu21} and those in the local universe \citep{kor13} with substituting dynamical/bulge mass for stellar mass.
The current stellar mass and BH mass obtained from the SED fitting and spectral fitting, respectively, for J1406 is shown in red star.
The open star represents the BH mass of J1406 using \cite{kor13} scaling relation with correction of redshift evolution following \cite{mer10}. 
    Blue arrow indicates growth of J1406 with constant Eddington ratio of $\lambdaedd = 3$ and radiation efficiency of $\eta=0.03$ for $10^{7}$~yr. Future location in the plane will be above the local scaling relation of \cite{kor13} (black line) by a factor of a few, which is close to the scaling relation of the high-$z$ massive quasars~\citep[blue dashed line;][]{pen20}.
}\label{fig:BH_growth}
\end{figure}

\subsection{Future SMBH Growth of J1406}

The super-Eddington phase is considered to be a crucial path for BH growth. We discuss how this phase will shape the direction of future BH growth for J1406.
The SED fitting result indicates that the BH accretion rate of $\dot{M}_\mathrm{BH}=(1-\eta) L_\mathrm{AGN,bol}/(\eta c^2)=0.13\times (1-\eta)/\eta\  \msun$~yr$^{-1}$ and $\mathrm{SFR}=230\msun$~yr$^{-1}$.  
This indicates that SMBH and host galaxy are growing with $\dot{M}_\mathrm{BH}/\mathrm{SFR} = 1/2000\times  (1-\eta)/\eta$. Assuming the fiducial value of $\eta=0.1$~\citep{sol82}, $\dot{M}_\mathrm{BH}/\mathrm{SFR}=5.1\times 10^{-3}$. This is comparable to the local scaling relation of $\mbh/\mstar \sim 3 \times 10^{-3}$~\citep{kor13}.

Figure~\ref{fig:BH_growth} shows the possible growth pathway of J1406 in the plane of $\mbh$ and $\mstar$ assuming the fiducial value of the standard disk of $\eta=0.1$. 
The red circle indicates the current $\mstar$ (based on the SED fitting) and $\mbh$ (based on spectral analysis) of J1406.
The growth of J1406 is only about 0.1 dex for both the black hole and the parent galaxy in the plane of Figure~\ref{fig:BH_growth}, when the current super-Eddington phase lasts about 10~Myr~\citep{kaw04,shi19,ina22}, with $\lambdaedd \sim 1$, $\eta=0.1$ and $\mathrm{SFR}=230~\msun$~yr$^{-1}$.
In contrast, super-Eddington phase has a radiatively inefficient disk~\citep[slim disk: e.g.,][for a review]{ina20}, 
resulting in efficient growth of BH with $\eta \leq 0.1$.
The radiation efficiency decreases into $\eta \simeq 0.03$~\citep{abr88,sad15} at the estimated Eddington ratio for J1406 of $\lambdaedd \simeq 3$ (but see \citealt{wat00} for $\eta = 0.1$ even at $\lambdaedd \simeq 3$). 
The blue arrow in Figure~\ref{fig:BH_growth}
shows the end point of the evolution after 10 Myr with the BH mass of $\sim 10^8\ \msun$. 
The resulting future location in the plane will move to the place above the local scaling relation of \cite{kor13} by a factor of a few and will be closer to the scaling relation of the high-$z$ massive quasars~\citep[blue dashed line;][]{pen20}. 
Large Eddington ratios are more easily to occur at higher redshifts, since galaxies are more gas-rich at higher-$z$~\citep{shi19}. 
Thus, J1406 and other radio-loud DOGs 
would be a key low-$z$ ($z<2$) analogous population of rapidly growing (over-massive) high-$z$ quasar as discussed in \cite{ina22} for super-Eddington accretion. 

\begin{figure}
    \centering
    \includegraphics[width=0.48\textwidth]{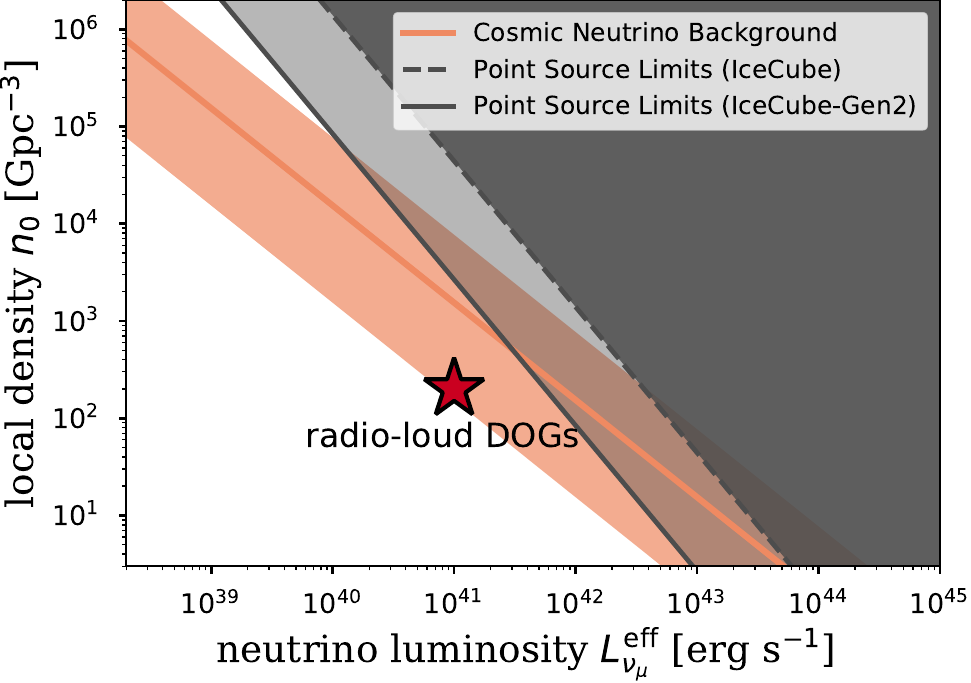}~
    \caption{Contribution to the diffuse neutrino flux (orange shaded region) of J1406-like radio-loud DOGs (red star) and IceCube detectability (black and grey shaded region). 
    This figure is made following Figure 3 of \cite{mur16}. 
    When the radio-loud DOGs have small reduction factor (e.g., $f_\mathrm{bol}=10$; red star), they can locate in above the required $L_\mathrm{\nu_\mu}^{\mathrm{eff}}$ and $n_0$ for redshift evolution of $\alpha\sim8$ (the bottom part of orange shaded region).  
}\label{fig:neutrino}
\end{figure}

\subsection{High-energy Neutrino Emission from Radio-loud DOGs}

Radio-loud AGNs are discussed as the source of high-energy cosmic rays (CRs; e.g., \citealt{tak90,rie22}). Cosmic rays are accelerated in the blazar zone, which is sub-pc away from the SMBH \citep[e.g.,][]{mur12} and/or extended jets, which are kpc away from the SMBH \citep[e.g.,][]{kim18b}. In both cases, the vast majority of CRs should escape from the jets, and they diffuse in the interstellar medium (ISM) of host galaxies. Radio-loud AGN in the local Universe are mostly elliptical galaxies, where the ISM gas density is so low ($n\sim0.01\rm~cm^{-3}$) that CRs cannot produce neutrinos efficiently via $pp$ interactions. On the other hand, radio-loud DOGs have a very dense gas ($n\sim100\rm~cm^{-3}$) in their ISM, which is suitable to produce neutrinos efficiently. Therefore, we discuss the possibility of detecting neutrinos from J1406 and the contribution of radio-loud DOGs to the cosmic neutrino background, whose origin has been in debates since the discovery by IceCube Collaboration~\citep[e.g.,][]{aar13}

We first estimate the neutrino flux from J1406. 
In radio-loud AGN, radio luminosity correlates with jet power~\citep{cav10}. Using the correlation, the jet power of J1406 is estimated to be $L_\mathrm{jet}\sim 10^{44}$~erg~s$^{-1}$. 
Then, the neutrino luminosity in the IceCube band is estimated to be \citep[e.g.,][]{kim19}
\begin{equation}
L_{\nu_\mu}^{\mathrm{eff}}=\frac{\epsilon_\mathrm{CR} f_{pp}  L_\mathrm{jet}}{6f_\mathrm{bol}} \sim 2\times10^{41}{\rm~erg~s^{-1}} \epsilon_\mathrm{CR,-1}f_{pp}f_\mathrm{bol,1}^{-1},
\end{equation}
where $\epsilon_\mathrm{CR}\sim 0.1$ is the CR production rate, $f_{pp}$ is the efficiency of pion production,  $f_\mathrm{bol}$ is the bolometric correction factor, and we use notation of $Q_x = Q/10^{x}$.
The bolometric correction factor $f_\mathrm{bol}$ is typically 10--20 for the diffusive shock acceleration model with the canonical spectral index of 2 \citep[e.g.,][]{kim22}, but it can be as low as a few, if the CR has a hard spectrum (e.g., stochastic acceleration by turbulence: \citealt{kim15}) and a large amount of flux is converted to the IceCube sensitivity range. 
This neutrino luminosity $L_{\nu_\mu}^{\mathrm{eff}}\sim10^{41.3}$~erg~s$^{-1}$ corresponds to the neutrino fluence of 
$E_\nu^2\phi_{\nu_\mu} \sim 3\times10^{-4}$~GeV~cm$^{-2}$ for 10 years of time integration, which is still difficult to detect even with the future detector, IceCube-Gen2~\citep{aar21}. 

Next, we discuss the contribution of radio-loud DOGs to cosmic high-energy neutrino background, which is determined by the comoving number density, $n_0$, and the effective neutrino luminosity, $L_{\nu_\mu}^{\rm eff}$.
Considering the survey area of radio-loud DOGs study of $\sim 105$~deg$^2$ \citep{nob19},
one source detection at $z=0.2$ (this work) with the comoving volume within the reshift of $\sim2$~Gpc$^3$ implies that the comoving number density of J1406-like radio-loud DOGs is estimated to be $n_0 \sim 200$~Gpc$^{-3}$. 
Using the neutrino luminosity of J1406, $L_\mathrm{\nu_\mu}^{\mathrm{eff}}$, and the local number density, $n_0$, we obtain the diffuse neutrino flux \citep[e.g.,][]{mur13}.
\begin{eqnarray}
E_\nu^2\Phi_{\nu_\mu}&\approx&\frac{cf_zL_\mathrm{\nu_\mu}^{\mathrm{eff}} n_0}{4\pi H_\mathrm{0}}  \\
&\simeq&0.2\times 10^{-8}{\rm~GeV~cm^{-2}~s^{-1}~sr^{-1}}f_{z,0.5}f_\mathrm{bol,1}^{-1},\nonumber
\end{eqnarray}
where $f_z$ is the redshift evolution factor. 
This value is comparable to the observed IceCube data of $1.44\times 10^{-8}\ (E_\nu/100\  \mathrm{TeV})^{-2.37}$~GeV~cm$^{-2}$~s$^{-1}$~sr$^{-1}$ for $\gtrsim $~100~TeV~\citep{abb22}. 
Here, $f_z=3$ roughly corresponds to the evolution of AGN \citep{boy98}, i.e., $n_0\propto (1+z)^\alpha$ with $\alpha\simeq3$~\citep{mur16}.
We note that the current fiducial value of $\alpha=3$ is very conservative since $\alpha=8$--$9.5$ for DOGs or ULIRGs \citep{got11,mag11}. We obtain $f_z\sim30$ with such a strong redshift evolution. Therefore,
the actual contribution by radio-loud DOGs would be much stronger. 

Figure~\ref{fig:neutrino} shows the required neutrino luminosity $L_\mathrm{\nu_\mu}^{\mathrm{eff}}$ and the local number density $n_0$ to account for the full diffuse neutrino flux observed by IceCube (orange shaded region), which is made following~\cite{mur16}. 
We can see that the radio-loud DOGs may potentially provide a sufficient amount of neutrinos if $f_z\gtrsim10$ (the lower part of the orange-shaded region in Figure~\ref{fig:neutrino}).  
Future surveys will clarify the local density of the radio-loud DOGs, $n_0$, the redshift evolution, $\alpha$,  and the luminosity of the radio jets, $L_{\rm jet}$, with which we will better model the neutrino emission and understand their contribution to the high-energy Universe. 


\acknowledgments

We thank the anonymous referee for helpful comments that improved the clarity and quality of the manuscript.
This work is supported by Japan Society for the Promotion of Science (JSPS) KAKENHI (25K01043; K.~Ichikawa) and KAKENHI  (22K14028, 21H04487; S.S. Kimura). S.S.K. acknowledges support by the Tohoku Initiative for Fostering Global Researchers for Interdisciplinary Sciences (TI-FRIS) of MEXT’s Strategic Professional Development Program for Young Researchers. 
We also acknowledge support from the National Natural Science Foundation of China (12073003, 12150410307, 12003003, 11721303, 11991052, 11950410493), and the China Manned Space Project Nos. CMS-CSST-2021- A04 and CMS-CSST-2021-A06. 

This publication makes use of data products from the Wide-field Infrared Survey Explorer, which is a joint project of the University of California, Los Angeles, and the Jet Propulsion Laboratory/California Institute of Technology, funded by the National Aeronautics and Space Administration.

\bibliographystyle{aasjournal}
\bibliography{bibtex}

\begin{thebibliography}{}
\expandafter\ifx\csname natexlab\endcsname\relax\def\natexlab#1{#1}\fi
\providecommand{\url}[1]{\href{#1}{#1}}
\providecommand{\dodoi}[1]{doi:~\href{http://doi.org/#1}{\nolinkurl{#1}}}
\providecommand{\doeprint}[1]{\href{http://ascl.net/#1}{\nolinkurl{http://ascl.net/#1}}}
\providecommand{\doarXiv}[1]{\href{https://arxiv.org/abs/#1}{\nolinkurl{https://arxiv.org/abs/#1}}}

\bibitem[{{Aartsen} {et~al.}(2021){Aartsen}, {Abbasi}, {Ackermann}, {Adams}, {Aguilar}, {Ahlers}, {Ahrens}, {Alispach}, {Allison}, {Amin}, {Andeen}, {Anderson}, {Ansseau}, {Anton}, {Arg{\"u}elles}, {Arlen}, {Auffenberg}, {Axani}, {Bagherpour}, {Bai}, {Balagopal V}, {Barbano}, {Bartos}, {Bastian}, {Basu}, {Baum}, {Baur}, {Bay}, {Beatty}, {Becker}, {Tjus}, {BenZvi}, {Berley}, {Bernardini}, {Besson}, {Binder}, {Bindig}, {Blaufuss}, {Blot}, {Bohm}, {Bohmer}, {B{\"o}ser}, {Botner}, {B{\"o}ttcher}, {Bourbeau}, {Bourbeau}, {Bradascio}, {Braun}, {Bron}, {Brostean-Kaiser}, {Burgman}, {Burley}, {Buscher}, {Busse}, {Bustamante}, {Campana}, {Carnie-Bronca}, {Carver}, {Chen}, {Chen}, {Cheung}, {Chirkin}, {Choi}, {Clark}, {Clark}, {Classen}, {Coleman}, {Collin}, {Connolly}, {Conrad}, {Coppin}, {Correa}, {Cowen}, {Cross}, {Dave}, {Deaconu}, {De Clercq}, {DeLaunay}, {De Kockere}, {Dembinski}, {Deoskar}, {De Ridder}, {Desai}, {Desiati}, {de Vries}, {de Wasseige}, {de With}, {DeYoung}, {Dharani}, {Diaz}, {D{\'\i}az-V{\'e}lez},
  {Dujmovic}, {Dunkman}, {DuVernois}, {Dvorak}, {Ehrhardt}, {Eller}, {Engel}, {Evans}, {Evenson}, {Fahey}, {Farrag}, {Fazely}, {Felde}, {Fienberg}, {Filimonov}, {Finley}, {Fischer}, {Fox}, {Franckowiak}, {Friedman}, {Fritz}, {Gaisser}, {Gallagher}, {Ganster}, {Garcia-Fernandez}, {Garrappa}, {Gartner}, {Gerhard}, {Gernhaeuser}, {Ghadimi}, {Glaser}, {Glauch}, {Gl{\"u}senkamp}, {Goldschmidt}, {Gonzalez}, {Goswami}, {Grant}, {Gr{\'e}goire}, {Griffith}, {Griswold}, {G{\"u}nd{\"u}z}, {Haack}, {Hallgren}, {Halliday}, {Halve}, {Halzen}, {Hanson}, {Hanson}, {Hardin}, {Haugen}, {Haungs}, {Hauser}, {Hebecker}, {Heinen}, {Heix}, {Helbing}, {Hellauer}, {Henningsen}, {Hickford}, {Hignight}, {Hill}, {Hill}, {Hoffman}, {Hoffmann}, {Hoffmann}, {Hoinka}, {Hokanson-Fasig}, {Holzapfel}, {Hoshina}, {Huang}, {Huber}, {Huber}, {Huege}, {Hughes}, {Hultqvist}, {H{\"u}nnefeld}, {Hussain}, {In}, {Iovine}, {Ishihara}, {Jansson}, {Japaridze}, {Jeong}, {Jones}, {Jonske}, {Joppe}, {Kalekin}, {Kang}, {Kang}, {Kang}, {Kappes}, {Kappesser},
  {Karg}, {Karl}, {Karle}, {Katori}, {Katz}, {Kauer}, {Keivani}, {Kellermann}, {Kelley}, {Kheirandish}, {Kim}, {Kin}, {Kintscher}, {Kiryluk}, {Kittler}, {Kleifges}, {Klein}, {Koirala}, {Kolanoski}, {K{\"o}pke}, {Kopper}, {Kopper}, {Koskinen}, {Koundal}, {Kovacevich}, {Kowalski}, {Krauss}, {Krings}, {Kr{\"u}ckl}, {Kulacz}, {Kurahashi}, {Gualda}, {Lahmann}, {Lanfranchi}, {Larson}, {Latif}, {Lauber}, {Lazar}, {Leonard}, {Leszczy{\'n}ska}, {Li}, {Liu}, {Lohfink}, {LoSecco}, {Mariscal}, {Lu}, {Lucarelli}, {Ludwig}, {L{\"u}nemann}, {Luszczak}, {Lyu}, {Ma}, {Madsen}, {Maggi}, {Mahn}, {Makino}, {Mallik}, {Mancina}, {Mandalia}, {Mari{\c{s}}}, {Marka}, {Marka}, {Maruyama}, {Mase}, {Maunu}, {McNally}, {Meagher}, {Medina}, {Meier}, {Meighen-Berger}, {Merz}, {Meyers}, {Micallef}, {Mockler}, {Moment{\'e}}, {Montaruli}, {Moore}, {Morse}, {Moulai}, {Muth}, {Naab}, {Nagai}, {Nam}, {Nauman}, {Necker}, {Neer}, {Nelles}, {Nguyễn}, {Niederhausen}, {Nisa}, {Nowicki}, {Nygren}, {Oberla}, {Pollmann}, {Oehler}, {Olivas},
  {O'Sullivan}, {Pan}, {Pandya}, {Pankova}, {Papp}, {Park}, {Parker}, {Paudel}, {Peiffer}, {P{\'e}rez de los Heros}, {Petersen}, {Philippen}, {Pieloth}, {Pieper}, {Pinfold}, {Pizzuto}, {Plaisier}, {Plum}, {Popovych}, {Porcelli}, {Rodriguez}, {Price}, {Przybylski}, {Raab}, {Raissi}, {Rameez}, {Rauch}, {Rawlins}, {Rea}, {Rehman}, {Reimann}, {Renschler}, {Renzi}, {Resconi}, {Reusch}, {Rhode}, {Richman}, {Riedel}, {Riegel}, {Roberts}, {Robertson}, {Roellinghoff}, {Rongen}, {Rott}, {Ruhe}, {Ryckbosch}, {Cantu}, {Safa}, {Herrera}, {Sandrock}, {Sandroos}, {Sandstrom}, {Santander}, {Sarkar}, {Sarkar}, {Satalecka}, {Scharf}, {Schaufel}, {Schieler}, {Schlunder}, {Schmidt}, {Schneider}, {Schneider}, {Schr{\"o}der}, {Schumacher}, {Sclafani}, {Seckel}, {Seunarine}, {Shaevitz}, {Sharma}, {Shefali}, {Silva}, {Smith}, {Smithers}, {Snihur}, {Soedingrekso}, {Soldin}, {S{\"o}ldner-Rembold}, {Song}, {Southall}, {Spiczak}, {Spiering}, {Stachurska}, {Stamatikos}, {Stanev}, {Stein}, {Stettner}, {Steuer}, {Stezelberger}, {Stokstad},
  {Strotjohann}, {St{\"u}rwald}, {Stuttard}, {Sullivan}, {Taboada}, {Taketa}, {Tanaka}, {Tenholt}, {Ter-Antonyan}, {Terliuk}, {Tilav}, {Tollefson}, {Tomankova}, {T{\"o}nnis}, {Torres}, {Toscano}, {Tosi}, {Trettin}, {Tselengidou}, {Tung}, {Turcati}, {Turcotte}, {Turley}, {Twagirayezu}, {Ty}, {Unger}, {Elorrieta}, {Vandenbroucke}, {van Eijk}, {van Eijndhoven}, {Vannerom}, {van Santen}, {Veberic}, {Verpoest}, {Vieregg}, {Vraeghe}, {Walck}, {Watson}, {Weaver}, {Weindl}, {Weinstock}, {Weiss}, {Weldert}, {Welling}, {Wendt}, {Werthebach}, {Whitehorn}, {Wiebe}, {Wiebusch}, {Williams}, {Wissel}, {Wolf}, {Wood}, {Woschnagg}, {Wrede}, {Wren}, {Wulff}, {Xu}, {Xu}, {Yanez}, {Yoshida}, {Yuan}, {Zhang}, {Zierke}, \& {Z{\"o}cklein}}]{aar21}
{Aartsen}, M.~G., {Abbasi}, R., {Ackermann}, M., {et~al.} 2021, Journal of Physics G Nuclear Physics, 48, 060501, \dodoi{10.1088/1361-6471/abbd48}

\bibitem[{{Abbasi} {et~al.}(2022){Abbasi}, {Ackermann}, {Adams}, {Aguilar}, {Ahlers}, {Ahrens}, {Alameddine}, {Alispach}, {Alves}, {Amin}, {Andeen}, {Anderson}, {Anton}, {Arg{\"u}elles}, {Ashida}, {Axani}, {Bai}, {Balagopal V.}, {Barbano}, {Barwick}, {Bastian}, {Basu}, {Baur}, {Bay}, {Beatty}, {Becker}, {Tjus}, {Bellenghi}, {BenZvi}, {Berley}, {Bernardini}, {Besson}, {Binder}, {Bindig}, {Blaufuss}, {Blot}, {Boddenberg}, {Bontempo}, {Borowka}, {B{\"o}ser}, {Botner}, {B{\"o}ttcher}, {Bourbeau}, {Bradascio}, {Braun}, {Brinson}, {Bron}, {Brostean-Kaiser}, {Browne}, {Burgman}, {Burley}, {Busse}, {Campana}, {Carnie-Bronca}, {Chen}, {Chen}, {Chirkin}, {Choi}, {Clark}, {Clark}, {Classen}, {Coleman}, {Collin}, {Conrad}, {Coppin}, {Correa}, {Cowen}, {Cross}, {Dappen}, {Dave}, {De Clercq}, {DeLaunay}, {L{\'o}pez}, {Dembinski}, {Deoskar}, {Desai}, {Desiati}, {de Vries}, {de Wasseige}, {de With}, {DeYoung}, {Diaz}, {D{\'\i}az-V{\'e}lez}, {Dittmer}, {Dujmovic}, {Dunkman}, {DuVernois}, {Dvorak}, {Ehrhardt}, {Eller},
  {Engel}, {Erpenbeck}, {Evans}, {Evenson}, {Fan}, {Fazely}, {Feigl}, {Fiedlschuster}, {Fienberg}, {Filimonov}, {Finley}, {Fischer}, {Fox}, {Franckowiak}, {Friedman}, {Fritz}, {F{\"u}rst}, {Gaisser}, {Gallagher}, {Ganster}, {Garcia}, {Garrappa}, {Gerhardt}, {Ghadimi}, {Glaser}, {Glauch}, {Gl{\"u}senkamp}, {Gonzalez}, {Goswami}, {Grant}, {Gr{\'e}goire}, {Griswold}, {G{\"u}nther}, {Gutjahr}, {Haack}, {Hallgren}, {Halliday}, {Halve}, {Halzen}, {Minh}, {Hanson}, {Hardin}, {Harnisch}, {Haungs}, {Hebecker}, {Helbing}, {Henningsen}, {Hettinger}, {Hickford}, {Hignight}, {Hill}, {Hill}, {Hoffman}, {Hoffmann}, {Hokanson-Fasig}, {Hoshina}, {Huang}, {Huber}, {Huber}, {Hultqvist}, {H{\"u}nnefeld}, {Hussain}, {Hymon}, {In}, {Iovine}, {Ishihara}, {Jansson}, {Japaridze}, {Jeong}, {Jin}, {Jones}, {Kang}, {Kang}, {Kang}, {Kappes}, {Kappesser}, {Kardum}, {Karg}, {Karl}, {Karle}, {Katz}, {Kauer}, {Kellermann}, {Kelley}, {Kheirandish}, {Kin}, {Kintscher}, {Kiryluk}, {Klein}, {Koirala}, {Kolanoski}, {Kontrimas}, {K{\"o}pke},
  {Kopper}, {Kopper}, {Koskinen}, {Koundal}, {Kovacevich}, {Kowalski}, {Kozynets}, {Kun}, {Kurahashi}, {Lad}, {Gualda}, {Lanfranchi}, {Larson}, {Lauber}, {Lazar}, {Lee}, {Leonard}, {Leszczy{\'n}ska}, {Li}, {Lincetto}, {Liu}, {Liubarska}, {Lohfink}, {Mariscal}, {Lu}, {Lucarelli}, {Ludwig}, {Luszczak}, {Lyu}, {Ma}, {Madsen}, {Mahn}, {Makino}, {Mancina}, {Mari{\c{s}}}, {Martinez-Soler}, {Maruyama}, {Mase}, {McElroy}, {McNally}, {Mead}, {Meagher}, {Mechbal}, {Medina}, {Meier}, {Meighen-Berger}, {Micallef}, {Mockler}, {Montaruli}, {Moore}, {Morse}, {Moulai}, {Naab}, {Nagai}, {Naumann}, {Necker}, {Nguyễn}, {Niederhausen}, {Nisa}, {Nowicki}, {Pollmann}, {Oehler}, {Oeyen}, {Olivas}, {O'Sullivan}, {Pandya}, {Pankova}, {Park}, {Parker}, {Paudel}, {Paul}, {de los Heros}, {Peters}, {Peterson}, {Philippen}, {Pieper}, {Pittermann}, {Pizzuto}, {Plum}, {Popovych}, {Porcelli}, {Rodriguez}, {Price}, {Pries}, {Przybylski}, {Raab}, {Raissi}, {Rameez}, {Rawlins}, {Rea}, {Rehman}, {Reichherzer}, {Reimann}, {Renzi}, {Resconi},
  {Reusch}, {Rhode}, {Richman}, {Riedel}, {Roberts}, {Robertson}, {Roellinghoff}, {Rongen}, {Rott}, {Ruhe}, {Ryckbosch}, {Cantu}, {Safa}, {Saffer}, {Herrera}, {Sandrock}, {Sandroos}, {Santander}, {Sarkar}, {Sarkar}, {Satalecka}, {Schaufel}, {Schieler}, {Schindler}, {Schmidt}, {Schneider}, {Schneider}, {Schr{\"o}der}, {Schumacher}, {Schwefer}, {Sclafani}, {Seckel}, {Seunarine}, {Sharma}, {Shefali}, {Silva}, {Skrzypek}, {Smithers}, {Snihur}, {Soedingrekso}, {Soldin}, {Spannfellner}, {Spiczak}, {Spiering}, {Stachurska}, {Stamatikos}, {Stanev}, {Stein}, {Stettner}, {Steuer}, {Stezelberger}, {St{\"u}rwald}, {Stuttard}, {Sullivan}, {Taboada}, {Ter-Antonyan}, {Tilav}, {Tischbein}, {Tollefson}, {T{\"o}nnis}, {Toscano}, {Tosi}, {Trettin}, {Tselengidou}, {Tung}, {Turcati}, {Turcotte}, {Turley}, {Twagirayezu}, {Ty}, {Elorrieta}, {Valtonen-Mattila}, {Vandenbroucke}, {van Eijndhoven}, {Vannerom}, {van Santen}, {Verpoest}, {Walck}, {Watson}, {Weaver}, {Weigel}, {Weindl}, {Weiss}, {Weldert}, {Wendt}, {Werthebach},
  {Weyrauch}, {Whitehorn}, {Wiebusch}, {Williams}, {Wolf}, {Woschnagg}, {Wrede}, {Wulff}, {Xu}, {Yanez}, {Yoshida}, {Yu}, {Yuan}, {Zhang}, {Zhelnin}, \& {IceCube Collaboration}}]{abb22}
{Abbasi}, R., {Ackermann}, M., {Adams}, J., {et~al.} 2022, \apj, 928, 50, \dodoi{10.3847/1538-4357/ac4d29}

\bibitem[{{Abdurro'uf} {et~al.}(2022){Abdurro'uf}, {Accetta}, {Aerts}, {Silva Aguirre}, {Ahumada}, {Ajgaonkar}, {Filiz Ak}, {Alam}, {Allende Prieto}, {Almeida}, {Anders}, {Anderson}, {Andrews}, {Anguiano}, {Aquino-Ort{\'\i}z}, {Arag{\'o}n-Salamanca}, {Argudo-Fern{\'a}ndez}, {Ata}, {Aubert}, {Avila-Reese}, {Badenes}, {Barb{\'a}}, {Barger}, {Barrera-Ballesteros}, {Beaton}, {Beers}, {Belfiore}, {Bender}, {Bernardi}, {Bershady}, {Beutler}, {Bidin}, {Bird}, {Bizyaev}, {Blanc}, {Blanton}, {Boardman}, {Bolton}, {Boquien}, {Borissova}, {Bovy}, {Brandt}, {Brown}, {Brownstein}, {Brusa}, {Buchner}, {Bundy}, {Burchett}, {Bureau}, {Burgasser}, {Cabang}, {Campbell}, {Cappellari}, {Carlberg}, {Wanderley}, {Carrera}, {Cash}, {Chen}, {Chen}, {Cherinka}, {Chiappini}, {Choi}, {Chojnowski}, {Chung}, {Clerc}, {Cohen}, {Comerford}, {Comparat}, {da Costa}, {Covey}, {Crane}, {Cruz-Gonzalez}, {Culhane}, {Cunha}, {Dai}, {Damke}, {Darling}, {Davidson}, {Davies}, {Dawson}, {De Lee}, {Diamond-Stanic}, {Cano-D{\'\i}az}, {S{\'a}nchez},
  {Donor}, {Duckworth}, {Dwelly}, {Eisenstein}, {Elsworth}, {Emsellem}, {Eracleous}, {Escoffier}, {Fan}, {Farr}, {Feng}, {Fern{\'a}ndez-Trincado}, {Feuillet}, {Filipp}, {Fillingham}, {Frinchaboy}, {Fromenteau}, {Galbany}, {Garc{\'\i}a}, {Garc{\'\i}a-Hern{\'a}ndez}, {Ge}, {Geisler}, {Gelfand}, {G{\'e}ron}, {Gibson}, {Goddy}, {Godoy-Rivera}, {Grabowski}, {Green}, {Greener}, {Grier}, {Griffith}, {Guo}, {Guy}, {Hadjara}, {Harding}, {Hasselquist}, {Hayes}, {Hearty}, {Hern{\'a}ndez}, {Hill}, {Hogg}, {Holtzman}, {Horta}, {Hsieh}, {Hsu}, {Hsu}, {Huber}, {Huertas-Company}, {Hutchinson}, {Hwang}, {Ibarra-Medel}, {Chitham}, {Ilha}, {Imig}, {Jaekle}, {Jayasinghe}, {Ji}, {Johnson}, {Jones}, {J{\"o}nsson}, {Katkov}, {Khalatyan}, {Kinemuchi}, {Kisku}, {Knapen}, {Kneib}, {Kollmeier}, {Kong}, {Kounkel}, {Kreckel}, {Krishnarao}, {Lacerna}, {Lane}, {Langgin}, {Lavender}, {Law}, {Lazarz}, {Leung}, {Leung}, {Lewis}, {Li}, {Li}, {Lian}, {Liang}, {Lin}, {Lin}, {Lin}, {Lintott}, {Long}, {Longa-Pe{\~n}a}, {L{\'o}pez-Cob{\'a}}, {Lu},
  {Lundgren}, {Luo}, {Mackereth}, {de la Macorra}, {Mahadevan}, {Majewski}, {Manchado}, {Mandeville}, {Maraston}, {Margalef-Bentabol}, {Masseron}, {Masters}, {Mathur}, {McDermid}, {Mckay}, {Merloni}, {Merrifield}, {Meszaros}, {Miglio}, {Di Mille}, {Minniti}, {Minsley}, {Monachesi}, {Moon}, {Mosser}, {Mulchaey}, {Muna}, {Mu{\~n}oz}, {Myers}, {Myers}, {Nadathur}, {Nair}, {Nandra}, {Neumann}, {Newman}, {Nidever}, {Nikakhtar}, {Nitschelm}, {O'Connell}, {Garma-Oehmichen}, {Luan Souza de Oliveira}, {Olney}, {Oravetz}, {Ortigoza-Urdaneta}, {Osorio}, {Otter}, {Pace}, {Padilla}, {Pan}, {Pan}, {Parikh}, {Parker}, {Peirani}, {Pe{\~n}a Ram{\'\i}rez}, {Penny}, {Percival}, {Perez-Fournon}, {Pinsonneault}, {Poidevin}, {Poovelil}, {Price-Whelan}, {B{\'a}rbara de Andrade Queiroz}, {Raddick}, {Ray}, {Rembold}, {Riddle}, {Riffel}, {Riffel}, {Rix}, {Robin}, {Rodr{\'\i}guez-Puebla}, {Roman-Lopes}, {Rom{\'a}n-Z{\'u}{\~n}iga}, {Rose}, {Ross}, {Rossi}, {Rubin}, {Salvato}, {S{\'a}nchez}, {S{\'a}nchez-Gallego}, {Sanderson}, {Santana
  Rojas}, {Sarceno}, {Sarmiento}, {Sayres}, {Sazonova}, {Schaefer}, {Schiavon}, {Schlegel}, {Schneider}, {Schultheis}, {Schwope}, {Serenelli}, {Serna}, {Shao}, {Shapiro}, {Sharma}, {Shen}, {Shetrone}, {Shu}, {Simon}, {Skrutskie}, {Smethurst}, {Smith}, {Sobeck}, {Spoo}, {Sprague}, {Stark}, {Stassun}, {Steinmetz}, {Stello}, {Stone-Martinez}, {Storchi-Bergmann}, {Stringfellow}, {Stutz}, {Su}, {Taghizadeh-Popp}, {Talbot}, {Tayar}, {Telles}, {Teske}, {Thakar}, {Theissen}, {Tkachenko}, {Thomas}, {Tojeiro}, {Hernandez Toledo}, {Troup}, {Trump}, {Trussler}, {Turner}, {Tuttle}, {Unda-Sanzana}, {V{\'a}zquez-Mata}, {Valentini}, {Valenzuela}, {Vargas-Gonz{\'a}lez}, {Vargas-Maga{\~n}a}, {Alfaro}, {Villanova}, {Vincenzo}, {Wake}, {Warfield}, {Washington}, {Weaver}, {Weijmans}, {Weinberg}, {Weiss}, {Westfall}, {Wild}, {Wilde}, {Wilson}, {Wilson}, {Wilson}, {Wolf}, {Wood-Vasey}, {Yan}, {Zamora}, {Zasowski}, {Zhang}, {Zhao}, {Zheng}, {Zheng}, \& {Zhu}}]{abd22}
{Abdurro'uf}, {Accetta}, K., {Aerts}, C., {et~al.} 2022, \apjs, 259, 35, \dodoi{10.3847/1538-4365/ac4414}

\bibitem[{{Abramowicz} {et~al.}(1988){Abramowicz}, {Czerny}, {Lasota}, \& {Szuszkiewicz}}]{abr88}
{Abramowicz}, M.~A., {Czerny}, B., {Lasota}, J.~P., \& {Szuszkiewicz}, E. 1988, \apj, 332, 646, \dodoi{10.1086/166683}

\bibitem[{{Abuter} {et~al.}(2024){Abuter}, {Allouche}, {Amorim}, {Bailet}, {Berdeu}, {Berger}, {Berio}, {Bigioli}, {Boebion}, {Bolzer}, {Bonnet}, {Bourdarot}, {Bourget}, {Brandner}, {Cao}, {Conzelmann}, {Comin}, {Cl{\'e}net}, {Courtney-Barrer}, {Davies}, {Defr{\`e}re}, {Delboulb{\'e}}, {Delplancke-Str{\"o}bele}, {Dembet}, {Dexter}, {de Zeeuw}, {Drescher}, {Eckart}, {{\'E}douard}, {Eisenhauer}, {Fabricius}, {Feuchtgruber}, {Finger}, {F{\"o}rster Schreiber}, {Garcia}, {Garcia Lopez}, {Gao}, {Gendron}, {Genzel}, {Gil}, {Gillessen}, {Gomes}, {Gont{\'e}}, {Gouvret}, {Guajardo}, {Guieu}, {Hackenberg}, {Haddad}, {Hartl}, {Haubois}, {Hau{\ss}mann}, {Hei{\ss}el}, {Henning}, {Hippler}, {H{\"o}nig}, {Horrobin}, {Hubin}, {Jacqmart}, {Jocou}, {Kaufer}, {Kervella}, {Kolb}, {Korhonen}, {Lacour}, {Lagarde}, {Lai}, {Lapeyr{\`e}re}, {Laugier}, {Le Bouquin}, {Leftley}, {L{\'e}na}, {Lewis}, {Liu}, {Lopez}, {Lutz}, {Magnard}, {Mang}, {Marcotto}, {Maurel}, {M{\'e}rand}, {Millour}, {More}, {Netzer}, {Nowacki}, {Nowak}, {Oberti},
  {Ott}, {Pallanca}, {Paumard}, {Perraut}, {Perrin}, {Petrov}, {Pfuhl}, {Pourr{\'e}}, {Rabien}, {Rau}, {Riquelme}, {Robbe-Dubois}, {Rochat}, {Salman}, {Sanchez-Bermudez}, {Santos}, {Scheithauer}, {Sch{\"o}ller}, {Schubert}, {Schuhler}, {Shangguan}, {Shchekaturov}, {Shimizu}, {Sevin}, {Soulez}, {Spang}, {Stadler}, {Sternberg}, {Straubmeier}, {Sturm}, {Sykes}, {Tacconi}, {Tristram}, {Vincent}, {von Fellenberg}, {Uysal}, {Widmann}, {Wieprecht}, {Wiezorrek}, {Woillez}, \& {Zins}}]{abu24}
{Abuter}, R., {Allouche}, F., {Amorim}, A., {et~al.} 2024, \nat, 627, 281, \dodoi{10.1038/s41586-024-07053-4}

\bibitem[{{Aihara} {et~al.}(2018{\natexlab{a}}){Aihara}, {Arimoto}, {Armstrong}, {Arnouts}, {Bahcall}, {Bickerton}, {Bosch}, {Bundy}, {Capak}, {Chan}, {Chiba}, {Coupon}, {Egami}, {Enoki}, {Finet}, {Fujimori}, {Fujimoto}, {Furusawa}, {Furusawa}, {Goto}, {Goulding}, {Greco}, {Greene}, {Gunn}, {Hamana}, {Harikane}, {Hashimoto}, {Hattori}, {Hayashi}, {Hayashi}, {He{\l}miniak}, {Higuchi}, {Hikage}, {Ho}, {Hsieh}, {Huang}, {Huang}, {Ikeda}, {Imanishi}, {Inoue}, {Iwasawa}, {Iwata}, {Jaelani}, {Jian}, {Kamata}, {Karoji}, {Kashikawa}, {Katayama}, {Kawanomoto}, {Kayo}, {Koda}, {Koike}, {Kojima}, {Komiyama}, {Konno}, {Koshida}, {Koyama}, {Kusakabe}, {Leauthaud}, {Lee}, {Lin}, {Lin}, {Lupton}, {Mandelbaum}, {Matsuoka}, {Medezinski}, {Mineo}, {Miyama}, {Miyatake}, {Miyazaki}, {Momose}, {More}, {More}, {Moritani}, {Moriya}, {Morokuma}, {Mukae}, {Murata}, {Murayama}, {Nagao}, {Nakata}, {Niida}, {Niikura}, {Nishizawa}, {Obuchi}, {Oguri}, {Oishi}, {Okabe}, {Okamoto}, {Okura}, {Ono}, {Onodera}, {Onoue}, {Osato}, {Ouchi},
  {Price}, {Pyo}, {Sako}, {Sawicki}, {Shibuya}, {Shimasaku}, {Shimono}, {Shirasaki}, {Silverman}, {Simet}, {Speagle}, {Spergel}, {Strauss}, {Sugahara}, {Sugiyama}, {Suto}, {Suyu}, {Suzuki}, {Tait}, {Takada}, {Takata}, {Tamura}, {Tanaka}, {Tanaka}, {Tanaka}, {Tanaka}, {Terai}, {Terashima}, {Toba}, {Tominaga}, {Toshikawa}, {Turner}, {Uchida}, {Uchiyama}, {Umetsu}, {Uraguchi}, {Urata}, {Usuda}, {Utsumi}, {Wang}, {Wang}, {Wong}, {Yabe}, {Yamada}, {Yamanoi}, {Yasuda}, {Yeh}, {Yonehara}, \& {Yuma}}]{aih18a}
{Aihara}, H., {Arimoto}, N., {Armstrong}, R., {et~al.} 2018{\natexlab{a}}, \pasj, 70, S4, \dodoi{10.1093/pasj/psx066}

\bibitem[{{Aihara} {et~al.}(2018{\natexlab{b}}){Aihara}, {Armstrong}, {Bickerton}, {Bosch}, {Coupon}, {Furusawa}, {Hayashi}, {Ikeda}, {Kamata}, {Karoji}, {Kawanomoto}, {Koike}, {Komiyama}, {Lang}, {Lupton}, {Mineo}, {Miyatake}, {Miyazaki}, {Morokuma}, {Obuchi}, {Oishi}, {Okura}, {Price}, {Takata}, {Tanaka}, {Tanaka}, {Tanaka}, {Uchida}, {Uraguchi}, {Utsumi}, {Wang}, {Yamada}, {Yamanoi}, {Yasuda}, {Arimoto}, {Chiba}, {Finet}, {Fujimori}, {Fujimoto}, {Furusawa}, {Goto}, {Goulding}, {Gunn}, {Harikane}, {Hattori}, {Hayashi}, {He{\l}miniak}, {Higuchi}, {Hikage}, {Ho}, {Hsieh}, {Huang}, {Huang}, {Imanishi}, {Iwata}, {Jaelani}, {Jian}, {Kashikawa}, {Katayama}, {Kojima}, {Konno}, {Koshida}, {Kusakabe}, {Leauthaud}, {Lee}, {Lin}, {Lin}, {Mandelbaum}, {Matsuoka}, {Medezinski}, {Miyama}, {Momose}, {More}, {More}, {Mukae}, {Murata}, {Murayama}, {Nagao}, {Nakata}, {Niida}, {Niikura}, {Nishizawa}, {Oguri}, {Okabe}, {Ono}, {Onodera}, {Onoue}, {Ouchi}, {Pyo}, {Shibuya}, {Shimasaku}, {Simet}, {Speagle}, {Spergel}, {Strauss},
  {Sugahara}, {Sugiyama}, {Suto}, {Suzuki}, {Tait}, {Takada}, {Terai}, {Toba}, {Turner}, {Uchiyama}, {Umetsu}, {Urata}, {Usuda}, {Yeh}, \& {Yuma}}]{aih18b}
{Aihara}, H., {Armstrong}, R., {Bickerton}, S., {et~al.} 2018{\natexlab{b}}, \pasj, 70, S8, \dodoi{10.1093/pasj/psx081}

\bibitem[{{Becker} {et~al.}(1995){Becker}, {White}, \& {Helfand}}]{bec95}
{Becker}, R.~H., {White}, R.~L., \& {Helfand}, D.~J. 1995, \apj, 450, 559, \dodoi{10.1086/176166}

\bibitem[{{Bell}(2003)}]{bel03}
{Bell}, E.~F. 2003, \apj, 586, 794, \dodoi{10.1086/367829}

\bibitem[{{Bertemes} {et~al.}(2024){Bertemes}, {Wylezalek}, {Rupke}, {Zakamska}, {Veilleux}, {Beckmann}, {Vayner}, {Sankar}, {Ishikawa}, {Diachenko}, {Liu}, {Chen}, {Seebeck}, {Lutz}, \& {Liu}}]{Ber24}
{Bertemes}, C., {Wylezalek}, D., {Rupke}, D. S.~N., {et~al.} 2024, arXiv e-prints, arXiv:2404.14475, \dodoi{10.48550/arXiv.2404.14475}

\bibitem[{{Bongiorno} {et~al.}(2012){Bongiorno}, {Merloni}, {Brusa}, {Magnelli}, {Salvato}, {Mignoli}, {Zamorani}, {Fiore}, {Rosario}, {Mainieri}, {Hao}, {Comastri}, {Vignali}, {Balestra}, {Bardelli}, {Berta}, {Civano}, {Kampczyk}, {Le Floc'h}, {Lusso}, {Lutz}, {Pozzetti}, {Pozzi}, {Riguccini}, {Shankar}, \& {Silverman}}]{bon12}
{Bongiorno}, A., {Merloni}, A., {Brusa}, M., {et~al.} 2012, \mnras, 427, 3103, \dodoi{10.1111/j.1365-2966.2012.22089.x}

\bibitem[{{Boquien} {et~al.}(2019){Boquien}, {Burgarella}, {Roehlly}, {Buat}, {Ciesla}, {Corre}, {Inoue}, \& {Salas}}]{boq19}
{Boquien}, M., {Burgarella}, D., {Roehlly}, Y., {et~al.} 2019, \aap, 622, A103, \dodoi{10.1051/0004-6361/201834156}

\bibitem[{{Boyle} \& {Terlevich}(1998)}]{boy98}
{Boyle}, B.~J., \& {Terlevich}, R.~J. 1998, \mnras, 293, L49, \dodoi{10.1046/j.1365-8711.1998.01264.x}

\bibitem[{{Bruzual} \& {Charlot}(2003)}]{bru03}
{Bruzual}, G., \& {Charlot}, S. 2003, \mnras, 344, 1000, \dodoi{10.1046/j.1365-8711.2003.06897.x}

\bibitem[{{Burtscher} {et~al.}(2013){Burtscher}, {Meisenheimer}, {Tristram}, {Jaffe}, {H{\"o}nig}, {Davies}, {Kishimoto}, {Pott}, {R{\"o}ttgering}, {Schartmann}, {Weigelt}, \& {Wolf}}]{bur13}
{Burtscher}, L., {Meisenheimer}, K., {Tristram}, K.~R.~W., {et~al.} 2013, \aap, 558, A149, \dodoi{10.1051/0004-6361/201321890}

\bibitem[{{Calzetti} {et~al.}(2000){Calzetti}, {Armus}, {Bohlin}, {Kinney}, {Koornneef}, \& {Storchi-Bergmann}}]{cal00}
{Calzetti}, D., {Armus}, L., {Bohlin}, R.~C., {et~al.} 2000, \apj, 533, 682, \dodoi{10.1086/308692}

\bibitem[{{Cavagnolo} {et~al.}(2010){Cavagnolo}, {McNamara}, {Nulsen}, {Carilli}, {Jones}, \& {B{\^\i}rzan}}]{cav10}
{Cavagnolo}, K.~W., {McNamara}, B.~R., {Nulsen}, P.~E.~J., {et~al.} 2010, \apj, 720, 1066, \dodoi{10.1088/0004-637X/720/2/1066}

\bibitem[{{Chabrier}(2003)}]{cha03}
{Chabrier}, G. 2003, \pasp, 115, 763, \dodoi{10.1086/376392}

\bibitem[{{Chen} {et~al.}(2020){Chen}, {Akiyama}, {Ichikawa}, {Noda}, {Toba}, {Yamamura}, {Kawaguchi}, {Abdurro'uf}, \& {Kokubo}}]{che20}
{Chen}, X., {Akiyama}, M., {Ichikawa}, K., {et~al.} 2020, \apj, 900, 51, \dodoi{10.3847/1538-4357/aba599}

\bibitem[{{Ciesla} {et~al.}(2015){Ciesla}, {Charmandaris}, {Georgakakis}, {Bernhard}, {Mitchell}, {Buat}, {Elbaz}, {LeFloc'h}, {Lacey}, {Magdis}, \& {Xilouris}}]{cie15}
{Ciesla}, L., {Charmandaris}, V., {Georgakakis}, A., {et~al.} 2015, \aap, 576, A10, \dodoi{10.1051/0004-6361/201425252}

\bibitem[{{Dale} {et~al.}(2014){Dale}, {Helou}, {Magdis}, {Armus}, {D{\'\i}az-Santos}, \& {Shi}}]{dal14}
{Dale}, D.~A., {Helou}, G., {Magdis}, G.~E., {et~al.} 2014, \apj, 784, 83, \dodoi{10.1088/0004-637X/784/1/83}

\bibitem[{{Dey} {et~al.}(2008){Dey}, {Soifer}, {Desai}, {Brand}, {Le Floc'h}, {Brown}, {Jannuzi}, {Armus}, {Bussmann}, {Brodwin}, {Bian}, {Eisenhardt}, {Higdon}, {Weedman}, \& {Willner}}]{dey08}
{Dey}, A., {Soifer}, B.~T., {Desai}, V., {et~al.} 2008, \apj, 677, 943, \dodoi{10.1086/529516}

\bibitem[{{Dong} {et~al.}(2008){Dong}, {Wang}, {Wang}, {Yuan}, {Zhou}, {Dai}, \& {Zhang}}]{don08}
{Dong}, X., {Wang}, T., {Wang}, J., {et~al.} 2008, \mnras, 383, 581, \dodoi{10.1111/j.1365-2966.2007.12560.x}

\bibitem[{{Edge} {et~al.}(2013){Edge}, {Sutherland}, {Kuijken}, {Driver}, {McMahon}, {Eales}, \& {Emerson}}]{edg13}
{Edge}, A., {Sutherland}, W., {Kuijken}, K., {et~al.} 2013, The Messenger, 154, 32

\bibitem[{{Eilers} {et~al.}(2020){Eilers}, {Hennawi}, {Decarli}, {Davies}, {Venemans}, {Walter}, {Ba{\~n}ados}, {Fan}, {Farina}, {Mazzucchelli}, {Novak}, {Schindler}, {Simcoe}, {Wang}, \& {Yang}}]{eil20}
{Eilers}, A.-C., {Hennawi}, J.~F., {Decarli}, R., {et~al.} 2020, \apj, 900, 37, \dodoi{10.3847/1538-4357/aba52e}

\bibitem[{{Fiore} {et~al.}(2017){Fiore}, {Feruglio}, {Shankar}, {Bischetti}, {Bongiorno}, {Brusa}, {Carniani}, {Cicone}, {Duras}, {Lamastra}, {Mainieri}, {Marconi}, {Menci}, {Maiolino}, {Piconcelli}, {Vietri}, \& {Zappacosta}}]{fio17}
{Fiore}, F., {Feruglio}, C., {Shankar}, F., {et~al.} 2017, \aap, 601, A143, \dodoi{10.1051/0004-6361/201629478}

\bibitem[{{Ganci} {et~al.}(2019){Ganci}, {Marziani}, {D'Onofrio}, {del Olmo}, {Bon}, {Bon}, \& {Negrete}}]{gan19}
{Ganci}, V., {Marziani}, P., {D'Onofrio}, M., {et~al.} 2019, \aap, 630, A110, \dodoi{10.1051/0004-6361/201936270}

\bibitem[{{Girardi} {et~al.}(2000){Girardi}, {Bressan}, {Bertelli}, \& {Chiosi}}]{gir00}
{Girardi}, L., {Bressan}, A., {Bertelli}, G., \& {Chiosi}, C. 2000, \aaps, 141, 371, \dodoi{10.1051/aas:2000126}

\bibitem[{{Gordon} {et~al.}(2021){Gordon}, {Boyce}, {O'Dea}, {Rudnick}, {Andernach}, {Vantyghem}, {Baum}, {Bui}, {Dionyssiou}, {Safi-Harb}, \& {Sander}}]{gor21}
{Gordon}, Y.~A., {Boyce}, M.~M., {O'Dea}, C.~P., {et~al.} 2021, \apjs, 255, 30, \dodoi{10.3847/1538-4365/ac05c0}

\bibitem[{{Goto} {et~al.}(2011){Goto}, {Arnouts}, {Inami}, {Matsuhara}, {Pearson}, {Takeuchi}, {Le Floc'h}, {Takagi}, {Wada}, {Nakagawa}, {Oyabu}, {Ishihara}, {Mok Lee}, {Jeong}, {Yamauchi}, {Serjeant}, {Sedgwick}, \& {Treister}}]{got11}
{Goto}, T., {Arnouts}, S., {Inami}, H., {et~al.} 2011, \mnras, 410, 573, \dodoi{10.1111/j.1365-2966.2010.17466.x}

\bibitem[{{Greene} \& {Ho}(2005)}]{gre05}
{Greene}, J.~E., \& {Ho}, L.~C. 2005, \apj, 630, 122, \dodoi{10.1086/431897}

\bibitem[{{Greene} {et~al.}(2020){Greene}, {Strader}, \& {Ho}}]{gre20}
{Greene}, J.~E., {Strader}, J., \& {Ho}, L.~C. 2020, \araa, 58, 257, \dodoi{10.1146/annurev-astro-032620-021835}

\bibitem[{{Grupe}(2004)}]{gru04}
{Grupe}, D. 2004, \aj, 127, 1799, \dodoi{10.1086/382516}

\bibitem[{{Heckman} {et~al.}(2004){Heckman}, {Kauffmann}, {Brinchmann}, {Charlot}, {Tremonti}, \& {White}}]{hec04}
{Heckman}, T.~M., {Kauffmann}, G., {Brinchmann}, J., {et~al.} 2004, \apj, 613, 109, \dodoi{10.1086/422872}

\bibitem[{{Helfand} {et~al.}(2015){Helfand}, {White}, \& {Becker}}]{hel15}
{Helfand}, D.~J., {White}, R.~L., \& {Becker}, R.~H. 2015, \apj, 801, 26, \dodoi{10.1088/0004-637X/801/1/26}

\bibitem[{{H{\"o}nig}(2019)}]{hon19}
{H{\"o}nig}, S.~F. 2019, \apj, 884, 171, \dodoi{10.3847/1538-4357/ab4591}

\bibitem[{{H{\"o}nig} {et~al.}(2012){H{\"o}nig}, {Kishimoto}, {Antonucci}, {Marconi}, {Prieto}, {Tristram}, \& {Weigelt}}]{hon12}
{H{\"o}nig}, S.~F., {Kishimoto}, M., {Antonucci}, R., {et~al.} 2012, \apj, 755, 149, \dodoi{10.1088/0004-637X/755/2/149}

\bibitem[{{H{\"o}nig} {et~al.}(2013){H{\"o}nig}, {Kishimoto}, {Tristram}, {Prieto}, {Gandhi}, {Asmus}, {Antonucci}, {Burtscher}, {Duschl}, \& {Weigelt}}]{hon13}
{H{\"o}nig}, S.~F., {Kishimoto}, M., {Tristram}, K.~R.~W., {et~al.} 2013, \apj, 771, 87, \dodoi{10.1088/0004-637X/771/2/87}

\bibitem[{{Hopkins} {et~al.}(2006){Hopkins}, {Hernquist}, {Cox}, {Di Matteo}, {Robertson}, \& {Springel}}]{hop06}
{Hopkins}, P.~F., {Hernquist}, L., {Cox}, T.~J., {et~al.} 2006, \apjs, 163, 1, \dodoi{10.1086/499298}

\bibitem[{{Hopkins} {et~al.}(2008){Hopkins}, {Hernquist}, {Cox}, \& {Kere{\v{s}}}}]{hop08}
{Hopkins}, P.~F., {Hernquist}, L., {Cox}, T.~J., \& {Kere{\v{s}}}, D. 2008, \apjs, 175, 356, \dodoi{10.1086/524362}

\bibitem[{{Hopkins} {et~al.}(2004){Hopkins}, {Strauss}, {Hall}, {Richards}, {Cooper}, {Schneider}, {Vanden Berk}, {Jester}, {Brinkmann}, \& {Szokoly}}]{hop04}
{Hopkins}, P.~F., {Strauss}, M.~A., {Hall}, P.~B., {et~al.} 2004, \aj, 128, 1112, \dodoi{10.1086/423291}

\bibitem[{{Hviding} {et~al.}(2022){Hviding}, {Hainline}, {Rieke}, {Juneau}, {Lyu}, \& {Pucha}}]{hvi22}
{Hviding}, R.~E., {Hainline}, K.~N., {Rieke}, M., {et~al.} 2022, \aj, 163, 224, \dodoi{10.3847/1538-3881/ac5e33}

\bibitem[{{Hwang} {et~al.}(2018){Hwang}, {Zakamska}, {Alexandroff}, {Hamann}, {Greene}, {Perrotta}, \& {Richards}}]{hwa18}
{Hwang}, H.-C., {Zakamska}, N.~L., {Alexandroff}, R.~M., {et~al.} 2018, \mnras, 477, 830, \dodoi{10.1093/mnras/sty742}

\bibitem[{{IceCube Collaboration}(2013)}]{aar13}
{IceCube Collaboration}. 2013, Science, 342, 1242856, \dodoi{10.1126/science.1242856}

\bibitem[{{IceCube Collaboration} {et~al.}(2018){IceCube Collaboration}, {Aartsen}, {Ackermann}, {Adams}, {Aguilar}, {Ahlers}, {Ahrens}, {Al Samarai}, {Altmann}, {Andeen}, {Anderson}, {Ansseau}, {Anton}, {Arg{\"u}elles}, {Auffenberg}, {Axani}, {Bagherpour}, {Bai}, {Barron}, {Barwick}, {Baum}, {Bay}, {Beatty}, {Becker Tjus}, {Becker}, {BenZvi}, {Berley}, {Bernardini}, {Besson}, {Binder}, {Bindig}, {Blaufuss}, {Blot}, {Bohm}, {B{\"o}rner}, {Bos}, {B{\"o}ser}, {Botner}, {Bourbeau}, {Bourbeau}, {Bradascio}, {Braun}, {Brenzke}, {Bretz}, {Bron}, {Brostean-Kaiser}, {Burgman}, {Busse}, {Carver}, {Cheung}, {Chirkin}, {Christov}, {Clark}, {Classen}, {Coenders}, {Collin}, {Conrad}, {Coppin}, {Correa}, {Cowen}, {Cross}, {Dave}, {Day}, {de Andr{\'e}}, {De Clercq}, {DeLaunay}, {Dembinski}, {De Ridder}, {Desiati}, {de Vries}, {de Wasseige}, {de With}, {DeYoung}, {D{\'\i}az-V{\'e}lez}, {di Lorenzo}, {Dujmovic}, {Dumm}, {Dunkman}, {Dvorak}, {Eberhardt}, {Ehrhardt}, {Eichmann}, {Eller}, {Evenson}, {Fahey}, {Fazely}, {Felde},
  {Filimonov}, {Finley}, {Flis}, {Franckowiak}, {Friedman}, {Fritz}, {Gaisser}, {Gallagher}, {Gerhardt}, {Ghorbani}, {Glauch}, {Gl{\"u}senkamp}, {Goldschmidt}, {Gonzalez}, {Grant}, {Griffith}, {Haack}, {Hallgren}, {Halzen}, {Hanson}, {Hebecker}, {Heereman}, {Helbing}, {Hellauer}, {Hickford}, {Hignight}, {Hill}, {Hoffman}, {Hoffmann}, {Hoinka}, {Hokanson-Fasig}, {Hoshina}, {Huang}, {Huber}, {Hultqvist}, {H{\"u}nnefeld}, {Hussain}, {In}, {Iovine}, {Ishihara}, {Jacobi}, {Japaridze}, {Jeong}, {Jero}, {Jones}, {Kalaczynski}, {Kang}, {Kappes}, {Kappesser}, {Karg}, {Karle}, {Katz}, {Kauer}, {Keivani}, {Kelley}, {Kheirandish}, {Kim}, {Kim}, {Kintscher}, {Kiryluk}, {Kittler}, {Klein}, {Koirala}, {Kolanoski}, {K{\"o}pke}, {Kopper}, {Kopper}, {Koschinsky}, {Koskinen}, {Kowalski}, {Krings}, {Kroll}, {Kr{\"u}ckl}, {Kunwar}, {Kurahashi}, {Kuwabara}, {Kyriacou}, {Labare}, {Lanfranchi}, {Larson}, {Lauber}, {Leonard}, {Lesiak-Bzdak}, {Leuermann}, {Liu}, {Lozano Mariscal}, {Lu}, {L{\"u}nemann}, {Luszczak}, {Madsen}, {Maggi},
  {Mahn}, {Mancina}, {Maruyama}, {Mase}, {Maunu}, {Meagher}, {Medici}, {Meier}, {Menne}, {Merino}, {Meures}, {Miarecki}, {Micallef}, {Moment{\'e}}, {Montaruli}, {Moore}, {Morse}, {Moulai}, {Nahnhauer}, {Nakarmi}, {Naumann}, {Neer}, {Niederhausen}, {Nowicki}, {Nygren}, {Obertacke Pollmann}, {Olivas}, {O'Murchadha}, {O'Sullivan}, {Palczewski}, {Pandya}, {Pankova}, {Peiffer}, {Pepper}, {P{\'e}rez de los Heros}, {Pieloth}, {Pinat}, {Plum}, {Price}, {Przybylski}, {Raab}, {R{\"a}del}, {Rameez}, {Rauch}, {Rawlins}, {Rea}, {Reimann}, {Relethford}, {Relich}, {Resconi}, {Rhode}, {Richman}, {Robertson}, {Rongen}, {Rott}, {Ruhe}, {Ryckbosch}, {Rysewyk}, {Safa}, {S{\"a}lzer}, {Sanchez Herrera}, {Sandrock}, {Sandroos}, {Santander}, {Sarkar}, {Sarkar}, {Satalecka}, {Schlunder}, {Schmidt}, {Schneider}, {Schoenen}, {Sch{\"o}neberg}, {Schumacher}, {Sclafani}, {Seckel}, {Seunarine}, {Soedingrekso}, {Soldin}, {Song}, {Spiczak}, {Spiering}, {Stachurska}, {Stamatikos}, {Stanev}, {Stasik}, {Stein}, {Stettner}, {Steuer},
  {Stezelberger}, {Stokstad}, {St{\"o}{\ss}l}, {Strotjohann}, {Stuttard}, {Sullivan}, {Sutherland}, {Taboada}, {Tatar}, {Tenholt}, {Ter-Antonyan}, {Terliuk}, {Tilav}, {Toale}, {Tobin}, {Toennis}, {Toscano}, {Tosi}, {Tselengidou}, {Tung}, {Turcati}, {Turley}, {Ty}, {Unger}, {Usner}, {Vandenbroucke}, {Van Driessche}, {van Eijk}, {van Eijndhoven}, {Vanheule}, {van Santen}, {Vogel}, {Vraeghe}, {Walck}, {Wallace}, {Wallraff}, {Wandler}, {Wandkowsky}, {Waza}, {Weaver}, {Weiss}, {Wendt}, {Werthebach}, {Westerhoff}, {Whelan}, {Whitehorn}, {Wiebe}, {Wiebusch}, {Wille}, {Williams}, {Wills}, {Wolf}, {Wood}, {Wood}, {Woschnagg}, {Xu}, {Xu}, {Xu}, {Yanez}, {Yodh}, {Yoshida}, {Yuan}, {Fermi-LAT Collaboration}, {Abdollahi}, {Ajello}, {Angioni}, {Baldini}, {Ballet}, {Barbiellini}, {Bastieri}, {Bechtol}, {Bellazzini}, {Berenji}, {Bissaldi}, {Blandford}, {Bonino}, {Bottacini}, {Bregeon}, {Bruel}, {Buehler}, {Burnett}, {Burns}, {Buson}, {Cameron}, {Caputo}, {Caraveo}, {Cavazzuti}, {Charles}, {Chen}, {Cheung}, {Chiang},
  {Chiaro}, {Ciprini}, {Cohen-Tanugi}, {Conrad}, {Costantin}, {Cutini}, {D'Ammando}, {de Palma}, {Digel}, {Di Lalla}, {Di Mauro}, {Di Venere}, {Dom{\'\i}nguez}, {Favuzzi}, {Franckowiak}, {Fukazawa}, {Funk}, {Fusco}, {Gargano}, {Gasparrini}, {Giglietto}, {Giomi}, {Giommi}, {Giordano}, {Giroletti}, {Glanzman}, {Green}, {Grenier}, {Grondin}, {Guiriec}, {Harding}, {Hayashida}, {Hays}, {Hewitt}, {Horan}, {J{\'o}hannesson}, {Kadler}, {Kensei}, {Kocevski}, {Krauss}, {Kreter}, {Kuss}, {La Mura}, {Larsson}, {Latronico}, {Lemoine-Goumard}, {Li}, {Longo}, {Loparco}, {Lovellette}, {Lubrano}, {Magill}, {Maldera}, {Malyshev}, {Manfreda}, {Mazziotta}, {McEnery}, {Meyer}, {Michelson}, {Mizuno}, {Monzani}, {Morselli}, {Moskalenko}, {Negro}, {Nuss}, {Ojha}, {Omodei}, {Orienti}, {Orlando}, {Palatiello}, {Paliya}, {Perkins}, {Persic}, {Pesce-Rollins}, {Piron}, {Porter}, {Principe}, {Rain{\`o}}, {Rando}, {Rani}, {Razzano}, {Razzaque}, {Reimer}, {Reimer}, {Renault-Tinacci}, {Ritz}, {Rochester}, {Saz Parkinson}, {Sgr{\`o}},
  {Siskind}, {Spandre}, {Spinelli}, {Suson}, {Tajima}, {Takahashi}, {Tanaka}, {Thayer}, {Thompson}, {Tibaldo}, {Torres}, {Torresi}, {Tosti}, {Troja}, {Valverde}, {Vianello}, {Vogel}, {Wood}, {Wood}, {Zaharijas}, {MAGIC Collaboration}, {Ahnen}, {Ansoldi}, {Antonelli}, {Arcaro}, {Baack}, {Babi{\'c}}, {Banerjee}, {Bangale}, {Barres de Almeida}, {Barrio}, {Becerra Gonz{\'a}lez}, {Bednarek}, {Bernardini}, {Berti}, {Bhattacharyya}, {Biland}, {Blanch}, {Bonnoli}, {Carosi}, {Carosi}, {Ceribella}, {Chatterjee}, {Colak}, {Colin}, {Colombo}, {Contreras}, {Cortina}, {Covino}, {Cumani}, {Da Vela}, {Dazzi}, {De Angelis}, {De Lotto}, {Delfino}, {Delgado}, {Di Pierro}, {Dom{\'\i}nguez}, {Dominis Prester}, {Dorner}, {Doro}, {Einecke}, {Elsaesser}, {Fallah Ramazani}, {Fern{\'a}ndez-Barral}, {Fidalgo}, {Foffano}, {Pfrang}, {Fonseca}, {Font}, {Franceschini}, {Fruck}, {Galindo}, {Gallozzi}, {Garc{\'\i}a L{\'o}pez}, {Garczarczyk}, {Gaug}, {Giammaria}, {Godinovi{\'c}}, {Gora}, {Guberman}, {Hadasch}, {Hahn}, {Hassan}, {Hayashida},
  {Herrera}, {Hose}, {Hrupec}, {Inoue}, {Ishio}, {Konno}, {Kubo}, {Kushida}, {Lelas}, {Lindfors}, {Lombardi}, {Longo}, {L{\'o}pez}, {Maggio}, {Majumdar}, {Makariev}, {Maneva}, {Manganaro}, {Mannheim}, {Maraschi}, {Mariotti}, {Mart{\'\i}nez}, {Masuda}, {Mazin}, {Minev}, {M}, {Mirzoyan}, {Moralejo}, {Moreno}, {Moretti}, {Nagayoshi}, {Neustroev}, {Niedzwiecki}, {Nievas Rosillo}, {Nigro}, {Nilsson}, {Ninci}, {Nishijima}, {Noda}, {Nogu{\'e}s}, {Paiano}, {Palacio}, {Paneque}, {Paoletti}, {Paredes}, {Pedaletti}, {Peresano}, {Persic}, {Prada Moroni}, {Prandini}, {Puljak}, {Rodriguez Garcia}, {Reichardt}, {Rhode}, {Rib{\'o}}, {Rico}, {Righi}, {Rugliancich}, {Saito}, {Satalecka}, {Schweizer}, {Sitarek}, {{\v{S}}nidaric {\textasciiacute}}, {Sobczynska}, {Stamerra}, {Strzys}, {Suri{\'c}}, {Takahashi}, {Tavecchio}, {Temnikov}, {Terzi{\'c}}, {Teshima}, {Torres-Alb{\`a}}, {Treves}, {Tsujimoto}, {Vanzo}, {Vazquez Acosta}, {Vovk}, {Ward}, {Will}, {S}, {Zaric {\textasciiacute}}, {AGILE Team}, {Lucarelli}, {Tavani}, {Piano},
  {Donnarumma}, {Pittori}, {Verrecchia}, {Barbiellini}, {Bulgarelli}, {Caraveo}, {Cattaneo}, {Colafrancesco}, {Costa}, {Di Cocco}, {Ferrari}, {Gianotti}, {Giuliani}, {Lipari}, {Mereghetti}, {Morselli}, {Pacciani}, {Paoletti}, {Parmiggiani}, {Pellizzoni}, {Picozza}, {Pilia}, {Rappoldi}, {Trois}, {Vercellone}, {Vittorini}, {ASAS-SN Team}, {Stanek}, {Franckowiak}, {Kochanek}, {Beacom}, {Thompson}, {Holoien}, {Dong}, {Prieto}, {Shappee}, {Holmbo}, {HAWC Collaboration}, {Abeysekara}, {Albert}, {Alfaro}, {Alvarez}, {Arceo}, {Arteaga-Vel{\'a}zquez}, {Avila Rojas}, {Ayala Solares}, {Becerril}, {Belmont-Moreno}, {Bernal}, {Caballero-Mora}, {Capistr{\'a}n}, {Carrami{\~n}ana}, {Casanova}, {Castillo}, {Cotti}, {Cotzomi}, {Couti{\~n}o de Le{\'o}n}, {De Le{\'o}n}, {De la Fuente}, {Diaz Hernandez}, {Dichiara}, {Dingus}, {DuVernois}, {D{\'\i}az-V{\'e}lez}, {Ellsworth}, {Engel}, {Fiorino}, {Fleischhack}, {Fraija}, {Garc{\'\i}a-Gonz{\'a}lez}, {Garfias}, {Gonz{\'a}lez Mu{\~n}oz}, {Gonz{\'a}lez}, {Goodman}, {Hampel-Arias},
  {Harding}, {Hernandez}, {Hona}, {Hueyotl-Zahuantitla}, {Hui}, {H{\"u}ntemeyer}, {Iriarte}, {Jardin-Blicq}, {Joshi}, {Kaufmann}, {Kunde}, {Lara}, {Lauer}, {Lee}, {Lennarz}, {Le{\'o}n Vargas}, {Linnemann}, {Longinotti}, {Luis-Raya}, {Luna-Garc{\'\i}a}, {Malone}, {Marinelli}, {Martinez}, {Martinez-Castellanos}, {Mart{\'\i}nez-Castro}, {Mart{\'\i}nez-Huerta}, {Matthews}, {Miranda-Romagnoli}, {Moreno}, {Mostaf{\'a}}, {Nayerhoda}, {Nellen}, {Newbold}, {Nisa}, {Noriega-Papaqui}, {Pelayo}, {Pretz}, {P{\'e}rez-P{\'e}rez}, {Ren}, {Rho}, {Rivi{\`e}re}, {Rosa-Gonz{\'a}lez}, {Rosenberg}, {Ruiz-Velasco}, {Ruiz-Velasco}, {Salesa Greus}, {Sandoval}, {Schneider}, {Schoorlemmer}, {Sinnis}, {Smith}, {Springer}, {Surajbali}, {Tibolla}, {Tollefson}, {Torres}, {Villase{\~n}or}, {Weisgarber}, {Werner}, {Yapici}, {Gaurang}, {Zepeda}, {Zhou}, {{\'A}lvarez}, {H.~E.~S.~S. Collaboration}, {Abdalla}, {Ang{\"u}ner}, {Armand}, {Backes}, {Becherini}, {Berge}, {B{\"o}ttcher}, {Boisson}, {Bolmont}, {Bonnefoy}, {Bordas}, {Brun},
  {B{\"u}chele}, {Bulik}, {Caroff}, {Carosi}, {Casanova}, {Cerruti}, {Chakraborty}, {Chandra}, {Chen}, {Colafrancesco}, {Davids}, {Deil}, {Devin}, {Djannati-Ata{\"\i}}, {Egberts}, {Emery}, {Eschbach}, {Fiasson}, {Fontaine}, {Funk}, {F{\"u}{\ss}ling}, {Gallant}, {Gat{\'e}}, {Giavitto}, {Glawion}, {Glicenstein}, {Gottschall}, {Grondin}, {Haupt}, {Henri}, {Hinton}, {Hoischen}, {Holch}, {Huber}, {Jamrozy}, {Jankowsky}, {Jankowsky}, {Jouvin}, {Jung-Richardt}, {Kerszberg}, {Kh{\'e}lifi}, {King}, {Klepser}, {Kluz {\textasciiacute}niak}, {Komin}, {Kraus}, {Lefaucheur}, {Lemi{\`e}re}, {Lemoine-Goumard}, {Lenain}, {Leser}, {Lohse}, {L{\'o}pez-Coto}, {Lorentz}, {Lypova}, {Marandon}, {Guillem Mart{\'\i}-Devesa}, {Maurin}, {Mitchell}, {Moderski}, {Mohamed}, {Mohrmann}, {Moulin}, {Murach}, {de Naurois}, {Niederwanger}, {Niemiec}, {Oakes}, {O'Brien}, {Ohm}, {Ostrowski}, {Oya}, {Panter}, {Parsons}, {Perennes}, {Piel}, {Pita}, {Poireau}, {Priyana Noel}, {Prokoph}, {P{\"u}hlhofer}, {Quirrenbach}, {Raab}, {Rauth}, {Renaud},
  {Rieger}, {Rinchiuso}, {Romoli}, {Rowell}, {Rudak}, {Sasaki}, {Sanchez}, {Schlickeiser}, {Sch{\"u}ssler}, {Schulz}, {Schwanke}, {Seglar-Arroyo}, {Shafi}, {Simoni}, {Sol}, {Stegmann}, {Steppa}, {Tavernier}, {Taylor}, {Tiziani}, {Trichard}, {Tsirou}, {van Eldik}, {van Rensburg}, {van Soelen}, {Veh}, {Vincent}, {Voisin}, {Wagner}, {Wagner}, {Wierzcholska}, {Zanin}, {Zdziarski}, {Zech}, {Ziegler}, {Zorn}, {{\.Z}ywucka}, {INTEGRAL Team}, {Savchenko}, {Ferrigno}, {Bazzano}, {Diehl}, {Kuulkers}, {Laurent}, {Mereghetti}, {Natalucci}, {Panessa}, {Rodi}, {Ubertini}, {Kanata}, Teams, {Morokuma}, {Ohta}, {Tanaka}, {Mori}, {Yamanaka}, {Kawabata}, {Utsumi}, {Nakaoka}, {Kawabata}, {Nagashima}, {Yoshida}, {Matsuoka}, {Itoh}, {Kapteyn Team}, {Keel}, {Liverpool Telescope Team}, {Copperwheat}, {Steele}, {Swift/NuSTAR Team}, {Cenko}, {Cowen}, {DeLaunay}, {Evans}, {Fox}, {Keivani}, {Kennea}, {Marshall}, {Osborne}, {Santander}, {Tohuvavohu}, {Turley}, {VERITAS Collaboration}, {Abeysekara}, {Archer}, {Benbow}, {Bird}, {Brill},
  {Brose}, {Buchovecky}, {Buckley}, {Bugaev}, {Christiansen}, {Connolly}, {Cui}, {Daniel}, {Errando}, {Falcone}, {Feng}, {Finley}, {Fortson}, {Furniss}, {Gueta}, {H{\"u}tten}, {Hervet}, {Hughes}, {Humensky}, {Johnson}, {Kaaret}, {Kar}, {Kelley-Hoskins}, {Kertzman}, {Kieda}, {Krause}, {Krennrich}, {Kumar}, {Lang}, {Lin}, {Maier}, {McArthur}, {Moriarty}, {Mukherjee}, {Nieto}, {O'Brien}, {Ong}, {Otte}, {Park}, {Petrashyk}, {Pohl}, {Popkow}, {Pueschel}, {Quinn}, {Ragan}, {Reynolds}, {Richards}, {Roache}, {Rulten}, {Sadeh}, {Santander}, {Scott}, {Sembroski}, {Shahinyan}, {Sushch}, {Tr{\'e}panier}, {Tyler}, {Vassiliev}, {Wakely}, {Weinstein}, {Wells}, {Wilcox}, {Wilhelm}, {Williams}, {Zitzer}, {VLA/B Team}, {Tetarenko}, {Kimball}, {Miller-Jones}, \& {Sivakoff}}]{ice18}
{IceCube Collaboration}, {Aartsen}, M.~G., {Ackermann}, M., {et~al.} 2018, Science, 361, eaat1378, \dodoi{10.1126/science.aat1378}

\bibitem[{{IceCube Collaboration} {et~al.}(2022){IceCube Collaboration}, {Abbasi}, {Ackermann}, {Adams}, {Aguilar}, {Ahlers}, {Ahrens}, {Alameddine}, {Alispach}, {Alves}, {Amin}, {Andeen}, {Anderson}, {Anton}, {Arg{\"u}elles}, {Ashida}, {Axani}, {Bai}, {Balagopal}, {Barbano}, {Barwick}, {Bastian}, {Basu}, {Baur}, {Bay}, {Beatty}, {Becker}, {Becker Tjus}, {Bellenghi}, {Benzvi}, {Berley}, {Bernardini}, {Besson}, {Binder}, {Bindig}, {Blaufuss}, {Blot}, {Boddenberg}, {Bontempo}, {Borowka}, {B{\"o}ser}, {Botner}, {B{\"o}ttcher}, {Bourbeau}, {Bradascio}, {Braun}, {Brinson}, {Bron}, {Brostean-Kaiser}, {Browne}, {Burgman}, {Burley}, {Busse}, {Campana}, {Carnie-Bronca}, {Chen}, {Chen}, {Chirkin}, {Choi}, {Clark}, {Clark}, {Classen}, {Coleman}, {Collin}, {Conrad}, {Coppin}, {Correa}, {Cowen}, {Cross}, {Dappen}, {Dave}, {de Clercq}, {Delaunay}, {Delgado L{\'o}pez}, {Dembinski}, {Deoskar}, {Desai}, {Desiati}, {de Vries}, {de Wasseige}, {de With}, {Deyoung}, {Diaz}, {D{\'\i}az-V{\'e}lez}, {Dittmer}, {Dujmovic}, {Dunkman},
  {Duvernois}, {Dvorak}, {Ehrhardt}, {Eller}, {Engel}, {Erpenbeck}, {Evans}, {Evenson}, {Fan}, {Fazely}, {Fedynitch}, {Feigl}, {Fiedlschuster}, {Fienberg}, {Filimonov}, {Finley}, {Fischer}, {Fox}, {Franckowiak}, {Friedman}, {Fritz}, {F{\"u}rst}, {Gaisser}, {Gallagher}, {Ganster}, {Garcia}, {Garrappa}, {Gerhardt}, {Ghadimi}, {Glaser}, {Glauch}, {Gl{\"u}senkamp}, {Goldschmidt}, {Gonzalez}, {Goswami}, {Grant}, {Gr{\'e}goire}, {Griswold}, {G{\"u}nther}, {Gutjahr}, {Haack}, {Hallgren}, {Halliday}, {Halve}, {Halzen}, {Hanson}, {Hardin}, {Harnisch}, {Haungs}, {Hebecker}, {Helbing}, {Henningsen}, {Hettinger}, {Hickford}, {Hignight}, {Hill}, {Hill}, {Hoffman}, {Hoffmann}, {Hokanson-Fasig}, {Hoshina}, {Huang}, {Huber}, {Huber}, {Hultqvist}, {H{\"u}nnefeld}, {Hussain}, {Hymon}, {in}, {Iovine}, {Ishihara}, {Jansson}, {Japaridze}, {Jeong}, {Jin}, {Jones}, {Kang}, {Kang}, {Kang}, {Kappes}, {Kappesser}, {Kardum}, {Karg}, {Karl}, {Karle}, {Katz}, {Kauer}, {Kellermann}, {Kelley}, {Kheirandish}, {Kin}, {Kintscher}, {Kiryluk},
  {Klein}, {Koirala}, {Kolanoski}, {Kontrimas}, {K{\"o}pke}, {Kopper}, {Kopper}, {Koskinen}, {Koundal}, {Kovacevich}, {Kowalski}, {Kozynets}, {Kun}, {Kurahashi}, {Lad}, {Lagunas Gualda}, {Lanfranchi}, {Larson}, {Lauber}, {Lazar}, {Lee}, {Leonard}, {Leszczy{\'n}ska}, {Li}, {Lincetto}, {Liu}, {Liubarska}, {Lohfink}, {Lozano Mariscal}, {Lu}, {Lucarelli}, {Ludwig}, {Luszczak}, {Lyu}, {Ma}, {Madsen}, {Mahn}, {Makino}, {Mancina}, {Mari{\c{s}}}, {Martinez-Soler}, {Maruyama}, {Mase}, {McElroy}, {McNally}, {Mead}, {Meagher}, {Mechbal}, {Medina}, {Meier}, {Meighen-Berger}, {Micallef}, {Mockler}, {Montaruli}, {Moore}, {Morse}, {Moulai}, {Naab}, {Nagai}, {Nahnhauer}, {Naumann}, {Necker}, {Nguyen}, {Niederhausen}, {Nisa}, {Nowicki}, {Nygren}, {Obertack}, {Pollmann}, {Oehler}, {Oeyen}, {Olivas}, {O'Sullivan}, {Pandya}, {Pankova}, {Park}, {Parker}, {Paudel}, {Paul}, {P{\'e}rez de Los Heros}, {Peters}, {Peterson}, {Philippen}, {Pieper}, {Pittermann}, {Pizzuto}, {Plum}, {Popovych}, {Porcelli}, {Prado Rodriguez}, {Price},
  {Pries}, {Przybylski}, {Rack-Helleis}, {Raissi}, {Rameez}, {Rawlins}, {Rea}, {Rehman}, {Reichherzer}, {Reimann}, {Renzi}, {Resconi}, {Reusch}, {Rhode}, {Richman}, {Riedel}, {Roberts}, {Robertson}, {Roellinghoff}, {Rongen}, {Rott}, {Ruhe}, {Ryckbosch}, {Rysewyk Cantu}, {Safa}, {Saffer}, {Sanchez Herrera}, {Sandrock}, {Sandroos}, {Santander}, {Sarkar}, {Sarkar}, {Satalecka}, {Schaufel}, {Schieler}, {Schindler}, {Schmidt}, {Schneider}, {Schneider}, {Schr{\"o}der}, {Schumacher}, {Schwefer}, {Sclafani}, {Seckel}, {Seunarine}, {Sharma}, {Shefali}, {Silva}, {Skrzypek}, {Smithers}, {Snihur}, {Soedingrekso}, {Soldin}, {Spannfellner}, {Spiczak}, {Spiering}, {Stachurska}, {Stamatikos}, {Stanev}, {Stein}, {Stettner}, {Steuer}, {Stezelberger}, {Stokstad}, {St{\"u}rwald}, {Stuttard}, {Sullivan}, {Taboada}, {Ter-Antonyan}, {Tilav}, {Tischbein}, {Tollefson}, {T{\"o}nnis}, {Toscano}, {Tosi}, {Trettin}, {Tselengidou}, {Tung}, {Turcati}, {Turcotte}, {Turley}, {Twagirayezu}, {Ty}, {Unland Elorrieta}, {Valtonen-Mattila},
  {Vandenbroucke}, {van Eijndhoven}, {Vannerom}, {van Santen}, {Verpoest}, {Walck}, {Watson}, {Weaver}, {Weigel}, {Weindl}, {Weiss}, {Weldert}, {Wendt}, {Werthebach}, {Weyrauch}, {Whitehorn}, {Wiebusch}, {Williams}, {Wolf}, {Woschnagg}, {Wrede}, {Wulff}, {Xu}, {Yanez}, {Yoshida}, {Yu}, {Yuan}, {Zhangan}, \& {Zhelnin}}]{Ice22}
{IceCube Collaboration}, {Abbasi}, R., {Ackermann}, M., {et~al.} 2022, Science, 378, 538, \dodoi{10.1126/science.abg3395}

\bibitem[{{Ichikawa} {et~al.}(2019{\natexlab{a}}){Ichikawa}, {Ueda}, {Bae}, {Kawamuro}, {Matsuoka}, {Toba}, \& {Shidatsu}}]{ich19b}
{Ichikawa}, K., {Ueda}, J., {Bae}, H.-J., {et~al.} 2019{\natexlab{a}}, \apj, 870, 65, \dodoi{10.3847/1538-4357/aaf233}

\bibitem[{{Ichikawa} {et~al.}(2016){Ichikawa}, {Ueda}, {Shidatsu}, {Kawamuro}, \& {Matsuoka}}]{ich16}
{Ichikawa}, K., {Ueda}, J., {Shidatsu}, M., {Kawamuro}, T., \& {Matsuoka}, K. 2016, \pasj, 68, 9, \dodoi{10.1093/pasj/psv112}

\bibitem[{{Ichikawa} {et~al.}(2012){Ichikawa}, {Ueda}, {Terashima}, {Oyabu}, {Gandhi}, {Matsuta}, \& {Nakagawa}}]{ich12}
{Ichikawa}, K., {Ueda}, Y., {Terashima}, Y., {et~al.} 2012, \apj, 754, 45, \dodoi{10.1088/0004-637X/754/1/45}

\bibitem[{{Ichikawa} {et~al.}(2015){Ichikawa}, {Packham}, {Ramos Almeida}, {Asensio Ramos}, {Alonso-Herrero}, {Gonz{\'a}lez-Mart{\'\i}n}, {Lopez-Rodriguez}, {Ueda}, {D{\'\i}az-Santos}, {Elitzur}, {H{\"o}nig}, {Imanishi}, {Levenson}, {Mason}, {Perlman}, \& {Alsip}}]{ich15}
{Ichikawa}, K., {Packham}, C., {Ramos Almeida}, C., {et~al.} 2015, \apj, 803, 57, \dodoi{10.1088/0004-637X/803/2/57}

\bibitem[{{Ichikawa} {et~al.}(2019{\natexlab{b}}){Ichikawa}, {Ricci}, {Ueda}, {Bauer}, {Kawamuro}, {Koss}, {Oh}, {Rosario}, {Shimizu}, {Stalevski}, {Fuller}, {Packham}, \& {Trakhtenbrot}}]{ich19}
{Ichikawa}, K., {Ricci}, C., {Ueda}, Y., {et~al.} 2019{\natexlab{b}}, \apj, 870, 31, \dodoi{10.3847/1538-4357/aaef8f}

\bibitem[{{Ichikawa} {et~al.}(2021){Ichikawa}, {Yamashita}, {Toba}, {Nagao}, {Inayoshi}, {Charisi}, {He}, {Wagner}, {Akiyama}, {Vijarnwannaluk}, {Chen}, {Kajisawa}, {Kawamuro}, {Lee}, {Matsuoka}, {Schramm}, {Suh}, {Tanaka}, {Uchiyama}, {Ueda}, {Pflugradt}, \& {Fukuchi}}]{ich21}
{Ichikawa}, K., {Yamashita}, T., {Toba}, Y., {et~al.} 2021, \apj, 921, 51, \dodoi{10.3847/1538-4357/ac1b26}

\bibitem[{{Ichikawa} {et~al.}(2023){Ichikawa}, {Yamashita}, {Merloni}, {Li}, {Liu}, {Salvato}, {Akiyama}, {Arcodia}, {Dwelly}, {Chen}, {Imanishi}, {Inayoshi}, {Kawaguchi}, {Kawamuro}, {Kokubo}, {Matsuoka}, {Nagao}, {Schramm}, {Suh}, {Tanaka}, {Toba}, \& {Ueda}}]{ich23}
{Ichikawa}, K., {Yamashita}, T., {Merloni}, A., {et~al.} 2023, \aap, 672, A171, \dodoi{10.1051/0004-6361/202244271}

\bibitem[{{Inayoshi}(2025)}]{ina25}
{Inayoshi}, K. 2025, arXiv e-prints, arXiv:2503.05537, \dodoi{10.48550/arXiv.2503.05537}

\bibitem[{{Inayoshi} {et~al.}(2022){Inayoshi}, {Nakatani}, {Toyouchi}, {Hosokawa}, {Kuiper}, \& {Onoue}}]{ina22}
{Inayoshi}, K., {Nakatani}, R., {Toyouchi}, D., {et~al.} 2022, \apj, 927, 237, \dodoi{10.3847/1538-4357/ac4751}

\bibitem[{{Inayoshi} {et~al.}(2020){Inayoshi}, {Visbal}, \& {Haiman}}]{ina20}
{Inayoshi}, K., {Visbal}, E., \& {Haiman}, Z. 2020, \araa, 58, 27, \dodoi{10.1146/annurev-astro-120419-014455}

\bibitem[{{Inoue} {et~al.}(2017){Inoue}, {Doi}, {Tanaka}, {Sikora}, \& {Madejski}}]{ino17}
{Inoue}, Y., {Doi}, A., {Tanaka}, Y.~T., {Sikora}, M., \& {Madejski}, G.~M. 2017, \apj, 840, 46, \dodoi{10.3847/1538-4357/aa6b57}

\bibitem[{{Intema} {et~al.}(2017){Intema}, {Jagannathan}, {Mooley}, \& {Frail}}]{int17}
{Intema}, H.~T., {Jagannathan}, P., {Mooley}, K.~P., \& {Frail}, D.~A. 2017, \aap, 598, A78, \dodoi{10.1051/0004-6361/201628536}

\bibitem[{{Izumi} {et~al.}(2021){Izumi}, {Matsuoka}, {Fujimoto}, {Onoue}, {Strauss}, {Umehata}, {Imanishi}, {Kohno}, {Kawaguchi}, {Kawamuro}, {Baba}, {Nagao}, {Toba}, {Inayoshi}, {Silverman}, {Inoue}, {Ikarashi}, {Iwasawa}, {Kashikawa}, {Hashimoto}, {Nakanishi}, {Ueda}, {Schramm}, {Lee}, \& {Suh}}]{izu21}
{Izumi}, T., {Matsuoka}, Y., {Fujimoto}, S., {et~al.} 2021, \apj, 914, 36, \dodoi{10.3847/1538-4357/abf6dc}

\bibitem[{{Kawada} {et~al.}(2007){Kawada}, {Baba}, {Barthel}, {Clements}, {Cohen}, {Doi}, {Figueredo}, {Fujiwara}, {Goto}, {Hasegawa}, {Hibi}, {Hirao}, {Hiromoto}, {Jeong}, {Kaneda}, {Kawai}, {Kawamura}, {Kester}, {Kii}, {Kobayashi}, {Kwon}, {Lee}, {Makiuti}, {Matsuo}, {Matsuura}, {M{\"u}ller}, {Murakami}, {Nagata}, {Nakagawa}, {Narita}, {Noda}, {Oh}, {Okada}, {Okuda}, {Oliver}, {Ootsubo}, {Pak}, {Park}, {Pearson}, {Rowan-Robinson}, {Saito}, {Salama}, {Sato}, {Savage}, {Serjeant}, {Shibai}, {Shirahata}, {Sohn}, {Suzuki}, {Takagi}, {Takahashi}, {Thomson}, {Usui}, {Verdugo}, {Watabe}, {White}, {Wang}, {Yamamura}, {Yamauchi}, \& {Yasuda}}]{kaw07}
{Kawada}, M., {Baba}, H., {Barthel}, P.~D., {et~al.} 2007, \pasj, 59, S389, \dodoi{10.1093/pasj/59.sp2.S389}

\bibitem[{{Kawaguchi} {et~al.}(2004){Kawaguchi}, {Aoki}, {Ohta}, \& {Collin}}]{kaw04}
{Kawaguchi}, T., {Aoki}, K., {Ohta}, K., \& {Collin}, S. 2004, \aap, 420, L23, \dodoi{10.1051/0004-6361:20040157}

\bibitem[{{Kelly} \& {Shen}(2013)}]{kel13}
{Kelly}, B.~C., \& {Shen}, Y. 2013, \apj, 764, 45, \dodoi{10.1088/0004-637X/764/1/45}

\bibitem[{{Kimura}(2022)}]{kim22}
{Kimura}, S.~S. 2022, arXiv e-prints, arXiv:2202.06480, \dodoi{10.48550/arXiv.2202.06480}

\bibitem[{{Kimura} {et~al.}(2019){Kimura}, {Murase}, \& {M{\'e}sz{\'a}ros}}]{kim19}
{Kimura}, S.~S., {Murase}, K., \& {M{\'e}sz{\'a}ros}, P. 2019, \prd, 100, 083014, \dodoi{10.1103/PhysRevD.100.083014}

\bibitem[{{Kimura} {et~al.}(2015){Kimura}, {Murase}, \& {Toma}}]{kim15}
{Kimura}, S.~S., {Murase}, K., \& {Toma}, K. 2015, \apj, 806, 159, \dodoi{10.1088/0004-637X/806/2/159}

\bibitem[{{Kimura} {et~al.}(2018){Kimura}, {Murase}, \& {Zhang}}]{kim18b}
{Kimura}, S.~S., {Murase}, K., \& {Zhang}, B.~T. 2018, \prd, 97, 023026, \dodoi{10.1103/PhysRevD.97.023026}

\bibitem[{{Komossa}(2008)}]{kom08}
{Komossa}, S. 2008, in Revista Mexicana de Astronomia y Astrofisica Conference Series, Vol.~32, Revista Mexicana de Astronomia y Astrofisica Conference Series, 86--92, \dodoi{10.48550/arXiv.0710.3326}

\bibitem[{{Komossa}(2018)}]{kom18}
{Komossa}, S. 2018, in Revisiting Narrow-Line Seyfert 1 Galaxies and their Place in the Universe, 15, \dodoi{10.22323/1.328.0015}

\bibitem[{{Kormendy} \& {Ho}(2013)}]{kor13}
{Kormendy}, J., \& {Ho}, L.~C. 2013, \araa, 51, 511, \dodoi{10.1146/annurev-astro-082708-101811}

\bibitem[{{Lacy} {et~al.}(2020){Lacy}, {Baum}, {Chandler}, {Chatterjee}, {Clarke}, {Deustua}, {English}, {Farnes}, {Gaensler}, {Gugliucci}, {Hallinan}, {Kent}, {Kimball}, {Law}, {Lazio}, {Marvil}, {Mao}, {Medlin}, {Mooley}, {Murphy}, {Myers}, {Osten}, {Richards}, {Rosolowsky}, {Rudnick}, {Schinzel}, {Sivakoff}, {Sjouwerman}, {Taylor}, {White}, {Wrobel}, {Andernach}, {Beasley}, {Berger}, {Bhatnager}, {Birkinshaw}, {Bower}, {Brandt}, {Brown}, {Burke-Spolaor}, {Butler}, {Comerford}, {Demorest}, {Fu}, {Giacintucci}, {Golap}, {G{\"u}th}, {Hales}, {Hiriart}, {Hodge}, {Horesh}, {Ivezi{\'c}}, {Jarvis}, {Kamble}, {Kassim}, {Liu}, {Loinard}, {Lyons}, {Masters}, {Mezcua}, {Moellenbrock}, {Mroczkowski}, {Nyland}, {O'Dea}, {O'Sullivan}, {Peters}, {Radford}, {Rao}, {Robnett}, {Salcido}, {Shen}, {Sobotka}, {Witz}, {Vaccari}, {van Weeren}, {Vargas}, {Williams}, \& {Yoon}}]{lac20}
{Lacy}, M., {Baum}, S.~A., {Chandler}, C.~J., {et~al.} 2020, \pasp, 132, 035001, \dodoi{10.1088/1538-3873/ab63eb}

\bibitem[{{Li} {et~al.}(2025){Li}, {Inayoshi}, {Chen}, {Ichikawa}, \& {Ho}}]{li25}
{Li}, Z., {Inayoshi}, K., {Chen}, K., {Ichikawa}, K., \& {Ho}, L.~C. 2025, \apj, 980, 36, \dodoi{10.3847/1538-4357/ada5fb}

\bibitem[{{Liu} {et~al.}(2021){Liu}, {Luo}, {Brandt}, {Brotherton}, {Gallagher}, {Ni}, {Shemmer}, \& {Timlin}}]{liu21}
{Liu}, H., {Luo}, B., {Brandt}, W.~N., {et~al.} 2021, \apj, 910, 103, \dodoi{10.3847/1538-4357/abe37f}

\bibitem[{{Lonsdale} {et~al.}(2015){Lonsdale}, {Lacy}, {Kimball}, {Blain}, {Whittle}, {Wilkes}, {Stern}, {Condon}, {Kim}, {Assef}, {Tsai}, {Efstathiou}, {Jones}, {Eisenhardt}, {Bridge}, {Wu}, {Lonsdale}, {Jones}, {Jarrett}, \& {Smith}}]{lon15}
{Lonsdale}, C.~J., {Lacy}, M., {Kimball}, A.~E., {et~al.} 2015, \apj, 813, 45, \dodoi{10.1088/0004-637X/813/1/45}

\bibitem[{{Lyu} \& {Rieke}(2018)}]{lyu18}
{Lyu}, J., \& {Rieke}, G.~H. 2018, \apj, 866, 92, \dodoi{10.3847/1538-4357/aae075}

\bibitem[{{Magnelli} {et~al.}(2011){Magnelli}, {Elbaz}, {Chary}, {Dickinson}, {Le Borgne}, {Frayer}, \& {Willmer}}]{mag11}
{Magnelli}, B., {Elbaz}, D., {Chary}, R.~R., {et~al.} 2011, \aap, 528, A35, \dodoi{10.1051/0004-6361/200913941}

\bibitem[{{Merloni} {et~al.}(2010){Merloni}, {Bongiorno}, {Bolzonella}, {Brusa}, {Civano}, {Comastri}, {Elvis}, {Fiore}, {Gilli}, {Hao}, {Jahnke}, {Koekemoer}, {Lusso}, {Mainieri}, {Mignoli}, {Miyaji}, {Renzini}, {Salvato}, {Silverman}, {Trump}, {Vignali}, {Zamorani}, {Capak}, {Lilly}, {Sanders}, {Taniguchi}, {Bardelli}, {Carollo}, {Caputi}, {Contini}, {Coppa}, {Cucciati}, {de la Torre}, {de Ravel}, {Franzetti}, {Garilli}, {Hasinger}, {Impey}, {Iovino}, {Iwasawa}, {Kampczyk}, {Kneib}, {Knobel}, {Kova{\v{c}}}, {Lamareille}, {Le Borgne}, {Le Brun}, {Le F{\`e}vre}, {Maier}, {Pello}, {Peng}, {Perez Montero}, {Ricciardelli}, {Scodeggio}, {Tanaka}, {Tasca}, {Tresse}, {Vergani}, \& {Zucca}}]{mer10}
{Merloni}, A., {Bongiorno}, A., {Bolzonella}, M., {et~al.} 2010, \apj, 708, 137, \dodoi{10.1088/0004-637X/708/1/137}

\bibitem[{{Miyazaki} {et~al.}(2018){Miyazaki}, {Komiyama}, {Kawanomoto}, {Doi}, {Furusawa}, {Hamana}, {Hayashi}, {Ikeda}, {Kamata}, {Karoji}, {Koike}, {Kurakami}, {Miyama}, {Morokuma}, {Nakata}, {Namikawa}, {Nakaya}, {Nariai}, {Obuchi}, {Oishi}, {Okada}, {Okura}, {Tait}, {Takata}, {Tanaka}, {Tanaka}, {Terai}, {Tomono}, {Uraguchi}, {Usuda}, {Utsumi}, {Yamada}, {Yamanoi}, {Aihara}, {Fujimori}, {Mineo}, {Miyatake}, {Oguri}, {Uchida}, {Tanaka}, {Yasuda}, {Takada}, {Murayama}, {Nishizawa}, {Sugiyama}, {Chiba}, {Futamase}, {Wang}, {Chen}, {Ho}, {Liaw}, {Chiu}, {Ho}, {Lai}, {Lee}, {Jeng}, {Iwamura}, {Armstrong}, {Bickerton}, {Bosch}, {Gunn}, {Lupton}, {Loomis}, {Price}, {Smith}, {Strauss}, {Turner}, {Suzuki}, {Miyazaki}, {Muramatsu}, {Yamamoto}, {Endo}, {Ezaki}, {Ito}, {Kawaguchi}, {Sofuku}, {Taniike}, {Akutsu}, {Dojo}, {Kasumi}, {Matsuda}, {Imoto}, {Miwa}, {Suzuki}, {Takeshi}, \& {Yokota}}]{miy18}
{Miyazaki}, S., {Komiyama}, Y., {Kawanomoto}, S., {et~al.} 2018, \pasj, 70, S1, \dodoi{10.1093/pasj/psx063}

\bibitem[{{Moster} {et~al.}(2013){Moster}, {Naab}, \& {White}}]{mos13}
{Moster}, B.~P., {Naab}, T., \& {White}, S. D.~M. 2013, \mnras, 428, 3121, \dodoi{10.1093/mnras/sts261}

\bibitem[{{Murase} {et~al.}(2013){Murase}, {Ahlers}, \& {Lacki}}]{mur13}
{Murase}, K., {Ahlers}, M., \& {Lacki}, B.~C. 2013, \prd, 88, 121301, \dodoi{10.1103/PhysRevD.88.121301}

\bibitem[{{Murase} {et~al.}(2012){Murase}, {Dermer}, {Takami}, \& {Migliori}}]{mur12}
{Murase}, K., {Dermer}, C.~D., {Takami}, H., \& {Migliori}, G. 2012, \apj, 749, 63, \dodoi{10.1088/0004-637X/749/1/63}

\bibitem[{{Murase} \& {Waxman}(2016)}]{mur16}
{Murase}, K., \& {Waxman}, E. 2016, \prd, 94, 103006, \dodoi{10.1103/PhysRevD.94.103006}

\bibitem[{{Nagai} {et~al.}(2006){Nagai}, {Inoue}, {Asada}, {Kameno}, \& {Doi}}]{nag06}
{Nagai}, H., {Inoue}, M., {Asada}, K., {Kameno}, S., \& {Doi}, A. 2006, \apj, 648, 148, \dodoi{10.1086/505793}

\bibitem[{{Noboriguchi} {et~al.}(2019){Noboriguchi}, {Nagao}, {Toba}, {Niida}, {Kajisawa}, {Onoue}, {Matsuoka}, {Yamashita}, {Chang}, {Kawaguchi}, {Komiyama}, {Nobuhara}, {Terashima}, \& {Ueda}}]{nob19}
{Noboriguchi}, A., {Nagao}, T., {Toba}, Y., {et~al.} 2019, \apj, 876, 132, \dodoi{10.3847/1538-4357/ab1754}

\bibitem[{{Noboriguchi} {et~al.}(2022){Noboriguchi}, {Nagao}, {Toba}, {Ichikawa}, {Kajisawa}, {Kato}, {Kawaguchi}, {Matsuhara}, {Matsuoka}, {Onishi}, {Onoue}, {Tamada}, {Terao}, {Terashima}, {Ueda}, \& {Yamashita}}]{nob22}
---. 2022, \apj, 941, 195, \dodoi{10.3847/1538-4357/aca403}

\bibitem[{{Noboriguchi} {et~al.}(2025){Noboriguchi}, {Ichikawa}, {Toba}, {Dwelly}, {Inayoshi}, {Kawaguchi}, {Liu}, {Terashima}, {Ueda}, {Akiyama}, {Brusa}, {Buchner}, {Kohno}, {Merloni}, {Nagao}, {Salvato}, {Suh}, \& {Urrutia}}]{nob25}
{Noboriguchi}, A., {Ichikawa}, K., {Toba}, Y., {et~al.} 2025, arXiv e-prints, arXiv:2504.09180, \dodoi{10.48550/arXiv.2504.09180}

\bibitem[{{O'Dea}(1998)}]{ode98}
{O'Dea}, C.~P. 1998, \pasp, 110, 493, \dodoi{10.1086/316162}

\bibitem[{{Ohsuga} {et~al.}(2009){Ohsuga}, {Mineshige}, {Mori}, \& {Kato}}]{ohs09}
{Ohsuga}, K., {Mineshige}, S., {Mori}, M., \& {Kato}, Y. 2009, \pasj, 61, L7, \dodoi{10.1093/pasj/61.3.L7}

\bibitem[{{Onaka} {et~al.}(2007){Onaka}, {Matsuhara}, {Wada}, {Fujishiro}, {Fujiwara}, {Ishigaki}, {Ishihara}, {Ita}, {Kataza}, {Kim}, {Matsumoto}, {Murakami}, {Ohyama}, {Oyabu}, {Sakon}, {Tanab{\'e}}, {Takagi}, {Uemizu}, {Ueno}, {Usui}, {Watarai}, {Cohen}, {Enya}, {Ootsubo}, {Pearson}, {Takeyama}, {Yamamuro}, \& {Ikeda}}]{ona07}
{Onaka}, T., {Matsuhara}, H., {Wada}, T., {et~al.} 2007, \pasj, 59, S401, \dodoi{10.1093/pasj/59.sp2.S401}

\bibitem[{{Patil} {et~al.}(2020){Patil}, {Nyland}, {Whittle}, {Lonsdale}, {Lacy}, {Lonsdale}, {Mukherjee}, {Trapp}, {Kimball}, {Lanz}, {Wilkes}, {Blain}, {Harwood}, {Efstathiou}, \& {Vlahakis}}]{pat20}
{Patil}, P., {Nyland}, K., {Whittle}, M., {et~al.} 2020, \apj, 896, 18, \dodoi{10.3847/1538-4357/ab9011}

\bibitem[{{Patil} {et~al.}(2022){Patil}, {Whittle}, {Nyland}, {Lonsdale}, {Lacy}, {Kimball}, {Lonsdale}, {Peters}, {Clarke}, {Efstathiou}, {Giacintucci}, {Kim}, {Lanz}, {Mukherjee}, \& {Polisensky}}]{pat22}
{Patil}, P., {Whittle}, M., {Nyland}, K., {et~al.} 2022, \apj, 934, 26, \dodoi{10.3847/1538-4357/ac71b0}

\bibitem[{{Pensabene} {et~al.}(2020){Pensabene}, {Carniani}, {Perna}, {Cresci}, {Decarli}, {Maiolino}, \& {Marconi}}]{pen20}
{Pensabene}, A., {Carniani}, S., {Perna}, M., {et~al.} 2020, \aap, 637, A84, \dodoi{10.1051/0004-6361/201936634}

\bibitem[{{Pilbratt} {et~al.}(2010){Pilbratt}, {Riedinger}, {Passvogel}, {Crone}, {Doyle}, {Gageur}, {Heras}, {Jewell}, {Metcalfe}, {Ott}, \& {Schmidt}}]{pil10}
{Pilbratt}, G.~L., {Riedinger}, J.~R., {Passvogel}, T., {et~al.} 2010, \aap, 518, L1, \dodoi{10.1051/0004-6361/201014759}

\bibitem[{{Prevot} {et~al.}(1984){Prevot}, {Lequeux}, {Maurice}, {Prevot}, \& {Rocca-Volmerange}}]{pre84}
{Prevot}, M.~L., {Lequeux}, J., {Maurice}, E., {Prevot}, L., \& {Rocca-Volmerange}, B. 1984, \aap, 132, 389

\bibitem[{{Ricci} {et~al.}(2023){Ricci}, {Ichikawa}, {Stalevski}, {Kawamuro}, {Yamada}, {Ueda}, {Mushotzky}, {Privon}, {Koss}, {Trakhtenbrot}, {Fabian}, {Ho}, {Asmus}, {Bauer}, {Chang}, {Gupta}, {Oh}, {Powell}, {Pfeifle}, {Rojas}, {Ricci}, {Temple}, {Toba}, {Tortosa}, {Treister}, {Harrison}, {Stern}, \& {Urry}}]{ric23}
{Ricci}, C., {Ichikawa}, K., {Stalevski}, M., {et~al.} 2023, \apj, 959, 27, \dodoi{10.3847/1538-4357/ad0733}

\bibitem[{{Rieger}(2022)}]{rie22}
{Rieger}, F.~M. 2022, Universe, 8, 607, \dodoi{10.3390/universe8110607}

\bibitem[{{Rieke} {et~al.}(2009){Rieke}, {Alonso-Herrero}, {Weiner}, {P{\'e}rez-Gonz{\'a}lez}, {Blaylock}, {Donley}, \& {Marcillac}}]{rie09}
{Rieke}, G.~H., {Alonso-Herrero}, A., {Weiner}, B.~J., {et~al.} 2009, \apj, 692, 556, \dodoi{10.1088/0004-637X/692/1/556}

\bibitem[{{Sadowski} \& {Narayan}(2015)}]{sad15}
{Sadowski}, A., \& {Narayan}, R. 2015, \mnras, 453, 3213, \dodoi{10.1093/mnras/stv1802}

\bibitem[{{Salvato} {et~al.}(2009){Salvato}, {Hasinger}, {Ilbert}, {Zamorani}, {Brusa}, {Scoville}, {Rau}, {Capak}, {Arnouts}, {Aussel}, {Bolzonella}, {Buongiorno}, {Cappelluti}, {Caputi}, {Civano}, {Cook}, {Elvis}, {Gilli}, {Jahnke}, {Kartaltepe}, {Impey}, {Lamareille}, {Le Floc'h}, {Lilly}, {Mainieri}, {McCarthy}, {McCracken}, {Mignoli}, {Mobasher}, {Murayama}, {Sasaki}, {Sanders}, {Schiminovich}, {Shioya}, {Shopbell}, {Silverman}, {Smol{\v{c}}i{\'c}}, {Surace}, {Taniguchi}, {Thompson}, {Trump}, {Urry}, \& {Zamojski}}]{sal09}
{Salvato}, M., {Hasinger}, G., {Ilbert}, O., {et~al.} 2009, \apj, 690, 1250, \dodoi{10.1088/0004-637X/690/2/1250}

\bibitem[{{Schlegel} {et~al.}(1998){Schlegel}, {Finkbeiner}, \& {Davis}}]{sch98}
{Schlegel}, D.~J., {Finkbeiner}, D.~P., \& {Davis}, M. 1998, \apj, 500, 525, \dodoi{10.1086/305772}

\bibitem[{{Shimwell} {et~al.}(2022){Shimwell}, {Hardcastle}, {Tasse}, {Best}, {R{\"o}ttgering}, {Williams}, {Botteon}, {Drabent}, {Mechev}, {Shulevski}, {van Weeren}, {Bester}, {Br{\"u}ggen}, {Brunetti}, {Callingham}, {Chy{\.z}y}, {Conway}, {Dijkema}, {Duncan}, {de Gasperin}, {Hale}, {Haverkorn}, {Hugo}, {Jackson}, {Mevius}, {Miley}, {Morabito}, {Morganti}, {Offringa}, {Oonk}, {Rafferty}, {Sabater}, {Smith}, {Schwarz}, {Smirnov}, {O'Sullivan}, {Vedantham}, {White}, {Albert}, {Alegre}, {Asabere}, {Bacon}, {Bonafede}, {Bonnassieux}, {Brienza}, {Bilicki}, {Bonato}, {Calistro Rivera}, {Cassano}, {Cochrane}, {Croston}, {Cuciti}, {Dallacasa}, {Danezi}, {Dettmar}, {Di Gennaro}, {Edler}, {En{\ss}lin}, {Emig}, {Franzen}, {Garc{\'\i}a-Vergara}, {Grange}, {G{\"u}rkan}, {Hajduk}, {Heald}, {Heesen}, {Hoang}, {Hoeft}, {Horellou}, {Iacobelli}, {Jamrozy}, {Jeli{\'c}}, {Kondapally}, {Kukreti}, {Kunert-Bajraszewska}, {Magliocchetti}, {Mahatma}, {Ma{\l}ek}, {Mandal}, {Massaro}, {Meyer-Zhao}, {Mingo}, {Mostert}, {Nair},
  {Nakoneczny}, {Nikiel-Wroczy{\'n}ski}, {Orr{\'u}}, {Pajdosz-{\'S}mierciak}, {Pasini}, {Prandoni}, {van Piggelen}, {Rajpurohit}, {Retana-Montenegro}, {Riseley}, {Rowlinson}, {Saxena}, {Schrijvers}, {Sweijen}, {Siewert}, {Timmerman}, {Vaccari}, {Vink}, {West}, {Wo{\l}owska}, {Zhang}, \& {Zheng}}]{shi22}
{Shimwell}, T.~W., {Hardcastle}, M.~J., {Tasse}, C., {et~al.} 2022, \aap, 659, A1, \dodoi{10.1051/0004-6361/202142484}

\bibitem[{{Shirakata} {et~al.}(2020){Shirakata}, {Kawaguchi}, {Okamoto}, {Nagashima}, \& {Oogi}}]{shi20}
{Shirakata}, H., {Kawaguchi}, T., {Okamoto}, T., {Nagashima}, M., \& {Oogi}, T. 2020, \apj, 898, 63, \dodoi{10.3847/1538-4357/ab9949}

\bibitem[{{Shirakata} {et~al.}(2019){Shirakata}, {Kawaguchi}, {Oogi}, {Okamoto}, \& {Nagashima}}]{shi19}
{Shirakata}, H., {Kawaguchi}, T., {Oogi}, T., {Okamoto}, T., \& {Nagashima}, M. 2019, \mnras, 487, 409, \dodoi{10.1093/mnras/stz1282}

\bibitem[{{Smith} {et~al.}(2017){Smith}, {Ibar}, {Maddox}, {Valiante}, {Dunne}, {Eales}, {Dye}, {Furlanetto}, {Bourne}, {Cigan}, {Ivison}, {Gomez}, {Smith}, \& {Viaene}}]{smi17}
{Smith}, M. W.~L., {Ibar}, E., {Maddox}, S.~J., {et~al.} 2017, \apjs, 233, 26, \dodoi{10.3847/1538-4365/aa9b35}

\bibitem[{{Soltan}(1982)}]{sol82}
{Soltan}, A. 1982, \mnras, 200, 115, \dodoi{10.1093/mnras/200.1.115}

\bibitem[{{Stalevski} {et~al.}(2012){Stalevski}, {Fritz}, {Baes}, {Nakos}, \& {Popovi{\'c}}}]{sta12}
{Stalevski}, M., {Fritz}, J., {Baes}, M., {Nakos}, T., \& {Popovi{\'c}}, L.~{\v{C}}. 2012, \mnras, 420, 2756, \dodoi{10.1111/j.1365-2966.2011.19775.x}

\bibitem[{{Stalevski} {et~al.}(2016){Stalevski}, {Ricci}, {Ueda}, {Lira}, {Fritz}, \& {Baes}}]{sta16}
{Stalevski}, M., {Ricci}, C., {Ueda}, Y., {et~al.} 2016, \mnras, 458, 2288, \dodoi{10.1093/mnras/stw444}

\bibitem[{{Sun} {et~al.}(2018){Sun}, {Greene}, {Zakamska}, {Goulding}, {Strauss}, {Huang}, {Johnson}, {Kawaguchi}, {Matsuoka}, {Marsteller}, {Nagao}, \& {Toba}}]{sun18}
{Sun}, A.-L., {Greene}, J.~E., {Zakamska}, N.~L., {et~al.} 2018, \mnras, 480, 2302, \dodoi{10.1093/mnras/sty1394}

\bibitem[{{Takahara}(1990)}]{tak90}
{Takahara}, F. 1990, Progress of Theoretical Physics, 83, 1071, \dodoi{10.1143/PTP.83.1071}

\bibitem[{{Takeuchi} {et~al.}(2010){Takeuchi}, {Ohsuga}, \& {Mineshige}}]{tak10}
{Takeuchi}, S., {Ohsuga}, K., \& {Mineshige}, S. 2010, \pasj, 62, L43, \dodoi{10.1093/pasj/62.5.L43}

\bibitem[{{Tazaki} \& {Ichikawa}(2020)}]{taz20b}
{Tazaki}, R., \& {Ichikawa}, K. 2020, \apj, 892, 149, \dodoi{10.3847/1538-4357/ab72f6}

\bibitem[{{Tazaki} {et~al.}(2020){Tazaki}, {Ichikawa}, \& {Kokubo}}]{taz20a}
{Tazaki}, R., {Ichikawa}, K., \& {Kokubo}, M. 2020, \apj, 892, 84, \dodoi{10.3847/1538-4357/ab7822}

\bibitem[{{Tchekhovskoy} {et~al.}(2011){Tchekhovskoy}, {Narayan}, \& {McKinney}}]{tch11}
{Tchekhovskoy}, A., {Narayan}, R., \& {McKinney}, J.~C. 2011, \mnras, 418, L79, \dodoi{10.1111/j.1745-3933.2011.01147.x}

\bibitem[{{Toba} {et~al.}(2015){Toba}, {Nagao}, {Strauss}, {Aoki}, {Goto}, {Imanishi}, {Kawaguchi}, {Terashima}, {Ueda}, {Bosch}, {Bundy}, {Doi}, {Inami}, {Komiyama}, {Lupton}, {Matsuhara}, {Matsuoka}, {Miyazaki}, {Morokuma}, {Nakata}, {Oi}, {Onoue}, {Oyabu}, {Price}, {Tait}, {Takata}, {Tanaka}, {Terai}, {Turner}, {Uchida}, {Usuda}, {Utsumi}, {Yamada}, \& {Wang}}]{tob15}
{Toba}, Y., {Nagao}, T., {Strauss}, M.~A., {et~al.} 2015, \pasj, 67, 86, \dodoi{10.1093/pasj/psv057}

\bibitem[{{Toba} {et~al.}(2017){Toba}, {Nagao}, {Kajisawa}, {Oogi}, {Akiyama}, {Ikeda}, {Coupon}, {Strauss}, {Wang}, {Tanaka}, {Niida}, {Imanishi}, {Lee}, {Matsuhara}, {Matsuoka}, {Onoue}, {Terashima}, {Ueda}, {Harikane}, {Komiyama}, {Miyazaki}, {Noboriguchi}, \& {Usuda}}]{tob17_clustering}
{Toba}, Y., {Nagao}, T., {Kajisawa}, M., {et~al.} 2017, \apj, 835, 36, \dodoi{10.3847/1538-4357/835/1/36}

\bibitem[{{Tortosa} {et~al.}(2023){Tortosa}, {Ricci}, {Ho}, {Tombesi}, {Du}, {Inayoshi}, {Wang}, {Shangguan}, \& {Li}}]{tor23}
{Tortosa}, A., {Ricci}, C., {Ho}, L.~C., {et~al.} 2023, \mnras, 519, 6267, \dodoi{10.1093/mnras/stac3590}

\bibitem[{{Vanden Berk} {et~al.}(2001){Vanden Berk}, {Richards}, {Bauer}, {Strauss}, {Schneider}, {Heckman}, {York}, {Hall}, {Fan}, {Knapp}, {Anderson}, {Annis}, {Bahcall}, {Bernardi}, {Briggs}, {Brinkmann}, {Brunner}, {Burles}, {Carey}, {Castander}, {Connolly}, {Crocker}, {Csabai}, {Doi}, {Finkbeiner}, {Friedman}, {Frieman}, {Fukugita}, {Gunn}, {Hennessy}, {Ivezi{\'c}}, {Kent}, {Kunszt}, {Lamb}, {Leger}, {Long}, {Loveday}, {Lupton}, {Meiksin}, {Merelli}, {Munn}, {Newberg}, {Newcomb}, {Nichol}, {Owen}, {Pier}, {Pope}, {Rockosi}, {Schlegel}, {Siegmund}, {Smee}, {Snir}, {Stoughton}, {Stubbs}, {SubbaRao}, {Szalay}, {Szokoly}, {Tremonti}, {Uomoto}, {Waddell}, {Yanny}, \& {Zheng}}]{van01}
{Vanden Berk}, D.~E., {Richards}, G.~T., {Bauer}, A., {et~al.} 2001, \aj, 122, 549, \dodoi{10.1086/321167}

\bibitem[{{Vazdekis} {et~al.}(2016){Vazdekis}, {Koleva}, {Ricciardelli}, {R{\"o}ck}, \& {Falc{\'o}n-Barroso}}]{vaz16}
{Vazdekis}, A., {Koleva}, M., {Ricciardelli}, E., {R{\"o}ck}, B., \& {Falc{\'o}n-Barroso}, J. 2016, \mnras, 463, 3409, \dodoi{10.1093/mnras/stw2231}

\bibitem[{{Wang} {et~al.}(2021){Wang}, {Yang}, {Fan}, {Hennawi}, {Barth}, {Banados}, {Bian}, {Boutsia}, {Connor}, {Davies}, {Decarli}, {Eilers}, {Farina}, {Green}, {Jiang}, {Li}, {Mazzucchelli}, {Nanni}, {Schindler}, {Venemans}, {Walter}, {Wu}, \& {Yue}}]{wan21}
{Wang}, F., {Yang}, J., {Fan}, X., {et~al.} 2021, \apjl, 907, L1, \dodoi{10.3847/2041-8213/abd8c6}

\bibitem[{{Watarai} {et~al.}(2000){Watarai}, {Fukue}, {Takeuchi}, \& {Mineshige}}]{wat00}
{Watarai}, K.-y., {Fukue}, J., {Takeuchi}, M., \& {Mineshige}, S. 2000, \pasj, 52, 133, \dodoi{10.1093/pasj/52.1.133}

\bibitem[{{Wright} {et~al.}(2010){Wright}, {Eisenhardt}, {Mainzer}, {Ressler}, {Cutri}, {Jarrett}, {Kirkpatrick}, {Padgett}, {McMillan}, {Skrutskie}, {Stanford}, {Cohen}, {Walker}, {Mather}, {Leisawitz}, {Gautier}, {McLean}, {Benford}, {Lonsdale}, {Blain}, {Mendez}, {Irace}, {Duval}, {Liu}, {Royer}, {Heinrichsen}, {Howard}, {Shannon}, {Kendall}, {Walsh}, {Larsen}, {Cardon}, {Schick}, {Schwalm}, {Abid}, {Fabinsky}, {Naes}, \& {Tsai}}]{wri10}
{Wright}, E.~L., {Eisenhardt}, P. R.~M., {Mainzer}, A.~K., {et~al.} 2010, \aj, 140, 1868, \dodoi{10.1088/0004-6256/140/6/1868}

\bibitem[{{Yang} {et~al.}(2020){Yang}, {Boquien}, {Buat}, {Burgarella}, {Ciesla}, {Duras}, {Stalevski}, {Brandt}, \& {Papovich}}]{yan20}
{Yang}, G., {Boquien}, M., {Buat}, V., {et~al.} 2020, \mnras, 491, 740, \dodoi{10.1093/mnras/stz3001}

\bibitem[{{Yang} {et~al.}(2022){Yang}, {Boquien}, {Brandt}, {Buat}, {Burgarella}, {Ciesla}, {Lehmer}, {Ma{\l}ek}, {Mountrichas}, {Papovich}, {Pons}, {Stalevski}, {Theul{\'e}}, \& {Zhu}}]{yan22}
{Yang}, G., {Boquien}, M., {Brandt}, W.~N., {et~al.} 2022, \apj, 927, 192, \dodoi{10.3847/1538-4357/ac4971}

\bibitem[{{Yoshida} {et~al.}(2025){Yoshida}, {Nagao}, {Toba}, {Noboriguchi}, {Ichikawa}, {Hildebrandt}, {Yutani}, {Chambers}, {Iwamoto}, {Kobayashi}, {Oguri}, {Osato}, {Shibata}, \& {Zhong}}]{yos25}
{Yoshida}, T., {Nagao}, T., {Toba}, Y., {et~al.} 2025, arXiv e-prints, arXiv:2504.15023, \dodoi{10.48550/arXiv.2504.15023}

\bibitem[{{Zakamska} \& {Greene}(2014)}]{zak14}
{Zakamska}, N.~L., \& {Greene}, J.~E. 2014, \mnras, 442, 784, \dodoi{10.1093/mnras/stu842}

\bibitem[{{Zakamska} {et~al.}(2016){Zakamska}, {Hamann}, {P{\^a}ris}, {Brandt}, {Greene}, {Strauss}, {Villforth}, {Wylezalek}, {Alexandroff}, \& {Ross}}]{zak16}
{Zakamska}, N.~L., {Hamann}, F., {P{\^a}ris}, I., {et~al.} 2016, \mnras, 459, 3144, \dodoi{10.1093/mnras/stw718}

\end{thebibliography}



\end{document}